%% file: aaskaii_template.tex
\documentclass[a4paper,11pt]{article}
\usepackage{aaskaiid}
\usepackage{orcidlink}

\title{Tracing cosmic star formation history through radio continuum spectral energy distribution and non-thermal emission}
\ShortTitle{SFH through SED \& non-thermal emission}

\author[1]{Fangxia An\orcidlink{0000-0001-7943-0166}$^{\star}$}
\ShortName{An, Tabatabaei, Seymour et al.} 
\author[2,3,4]{Fatemeh S. Tabatabaei\orcidlink{0000-0002-0377-0970}$^{\star}$}
\author[5]{Nick Seymour\orcidlink{0000-0003-3506-5536}$^{\star}$}
\author[2]{Maryam Khademi\orcidlink{0000-0002-2865-0692}}
\author[6]{Zunli Yuan\orcidlink{0000-0001-6861-0022}}
\author[6]{Wenjie Wang\orcidlink{0009-0005-1617-2442}}
\author[5]{Joe A. Grundy\orcidlink{0000-0002-4440-8046}}
\author[7]{Mark T. Sargent\orcidlink{0000-0003-1033-9684}}
\author[8]{Tao Wang\orcidlink{0000-0002-2504-2421}}
\author[8]{Yuheng Zhang\orcidlink{0000-0001-5757-5719}}
\author[1,9]{Yinghe Zhao\orcidlink{0000-0002-9128-818X}}
\author[10]{Hiddo S.~B. Algera\orcidlink{0000-0002-4205-9567}}
\author[11]{Elias Brinks\orcidlink{0000-0002-7758-9699}}
\author[12,13,14]{Mattia Vaccari\orcidlink{0000-0002-6748-0577}}

\affiliation[1]{Yunnan Observatories, Chinese Academy of Sciences, Kunming 650216, People's Republic of China}
\emailAdd{anfangxia@ynao.ac.cn}
\affiliation[2]{School of Astronomy, Institute for Research in Fundamental Sciences (IPM), PO Box 19395-5531, Tehran, Iran}
\emailAdd{ftaba@ipm.ir/tabata@ph1.uni-koeln.de}
\affiliation[3]{I. Physik. Institut, University of Cologne, 50937 Cologne, Germany}
\affiliation[4]{Max-Planck-Institut f\"ur Astronomie, Department of Galaxies and Cosmology, K\"onigstuhl 17, 69117 Heidelberg, Germany}
\affiliation[5]{International Centre for Radio Astronomy Research, Curtin University, GPO Box U1987, Bentley, WA 6845, Australia}
\emailAdd{nick.seymour@curtin.edu.au}
\affiliation[6]{Department of Physics, School of Physics and Electronics, Hunan Normal University, Changsha 410081, People's Republic of China}
\affiliation[7]{Institute of Physics, Laboratory of Astrophysics, \'Ecole Polytechnique F\'ed\'erale de Lausanne (EPFL), Observatoire de Sauverny, Versoix CH-1290, Switzerland}
\affiliation[8]{School of Astronomy and Space Science, Nanjing University, Nanjing, Jiangsu 210093, People's Republic of China}
\affiliation[9]{State Key Laboratory of Radio Astronomy and Technology, National Astronomical Observatories, Chinese Academy of Sciences, Beijing 100101, People's Republic of China}
\affiliation[10]{Institute of Astronomy and Astrophysics, Academia Sinica, 11F of Astronomy-Mathematics Building, No.1, Sec. 4, Roosevelt Rd, Taipei 106319, Taiwan, R.O.C.}
\affiliation[11]{Centre for Astrophysics Research, University of Hertfordshire, College Lane, Hatfield AL10 9AB, UK}
\affiliation[12]{Inter-University Institute for Data Intensive Astronomy (IDIA)-Department of Astronomy, University of Cape Town, Private Bag X3, 7701 Rondebosch, Cape Town, South Africa}
\affiliation[13]{Department of Physics and Astronomy, University of the Western Cape, 7535 Bellville, Cape Town, South Africa}
\affiliation[14]{INAF - Istituto di Radioastronomia, via Gobetti 101, 40129 Bologna, Italy}
\affiliation[\textsuperscript{$\star$}]{Corresponding author}

\abstract{As a tracer of massive star formation unaffected by dust, the radio continuum emission  provides a unique window into the formation of the first stars and galaxies in the Universe.  Recent observations show that the integrated rest-frame mid-radio {($\sim$1-10\,GHz)} luminosity of galaxies serves as one of the most robust tracers of the star formation rate (SFR). These studies {further demonstrate that the synchrotron spectral index, and more generally the shape of the radio spectral energy distribution (SED), evolves with redshift as a consequence of the cosmic evolution of star formation activity.} These findings underscore the importance of deep multi-band radio continuum observations in calibrating the SFR of early galaxies and understanding the astrophysical processes governing their assembly and evolution over cosmic time.
This chapter presents recent progress in radio SFR calibrations for star-forming galaxies (SFGs) and reviews radio-continuum studies of the cosmic star formation history (SFH). We highlight the transformative potential of SKA AA4, whose broad frequency coverage and high sensitivity will enable well-constrained radio SEDs for SFGs across a wide redshift range. {Using mock SKA AA4 continuum surveys based on the deep multi-band and ultra-deep Tiers, we show that the broad frequency coverage of the former will provide improved estimates of rest-frame radio luminosities through radio SED fitting, thereby enabling better-constrained radio-based measurements of the cosmic SFH. The Ultra-deep Tier, reaching a sensitivity of $\sim0.05\,\mu$Jy\,beam$^{-1}$ at $\sim1.4$\,GHz, will detect SFGs with SFRs down to $\sim1\,M_{\odot}\,{\rm yr}^{-1}$ at $z\sim5$ and $\sim10\,M_{\odot}\,{\rm yr}^{-1}$ at $z\sim8$. Such deep samples will provide strong constraints on the radio luminosity function out to $z\sim5$, especially its faint end at high redshift. Together, the broad frequency coverage and sensitivity of SKA AA4 will enable} more robust radio-based measurements of the cosmic SFH at epochs that remain challenging for current facilities.

}

\begin{document}
\maketitle

\section{Introduction}\label{sec:introduction}
To understand the assembly and evolution of galaxies, it is important to know the build-up of stellar mass over the history of the Universe, i.e., the cosmic star formation history (SFH), which relies on observational techniques to measure both ongoing star-formation activity and the stellar
populations that are already in place. Optical/NIR observatories such as the HST and JWST have made major breakthroughs tracing old stars, with the most recent discovery of red and mature galaxies already encountered within what is known as the "dark ages" \citep[$z\simeq14$,][]{robert} which challenges the standard cosmological model. Measuring the ongoing star formation rate (SFR) is, however, not as straightforward because it occurs within dusty molecular clouds, and its radiated blue light is highly affected by dust obscuration. Therefore, it is crucial to have an effective way to derive SFRs for galaxies spanning a large range of look-back times in order to understand galaxy evolution leading back to initial conditions.

Over the past two to three decades, significant progress has been made in reconstructing the cosmic star formation history (SFH). Beginning in the 1990s, with the initial discovery that the comoving star formation rate density (SFRD, denoted as $\rho_{\rm SFR}$ and expressed in units of $M_\odot$\,yr$^{-1}$Mpc$^{-3}$) at $z\sim1$ was substantially higher than in the local universe, extensive ultraviolet (UV), optical, and infrared (IR) observations have established a coherent picture of the SFRD evolution over cosmic time. These studies indicate that the cosmic SFRD increases rapidly in the early Universe, peaks around $z\sim$2-3, and subsequently declines by approximately an order of magnitude to the present day, as comprehensively reviewed by \cite{Hopkins06} and \cite{Madau14}.

While broad consensus exists regarding the SFRD evolution up to $z\sim$\,2 across UV, optical, and IR tracers, systematic discrepancies are already evident between estimates derived from shorter wavelengths (e.g., UV/optical) and those obtained from longer wavelengths (e.g., radio and submillimeter). ALMA observations have demonstrated that UV/optical-based SFRD estimates overlook a substantial population of obscured, dusty, and highly star-forming galaxies, which become increasingly prevalent at higher redshifts \citep{Wang19, Bouwens20}. Dust-obscured star formation accounts for $>$\,50\% of the total SFRD at $z\sim$2 \citep{Casey14,Zavala21}, and recent deep JWST studies of ALMA-detected galaxies suggest that its contribution remains significant even beyond $z>4$ \citep{Sun25,Liu26}. Moreover, recent measurements from the most sensitive $\sim$1.4\,GHz radio continuum surveys with MeerKAT and from the wide-area low-frequency observations with LOFAR have revealed a systematic excess in radio-based SFRDs compared to those derived from UV–IR tracers \citep{Cochrane23, Matthews24}. Although the origin of this excess remains under debate, these results reinforce the view that current measurements of the cosmic SFH, even at $z<2$, based primarily on UV/optically selected galaxy samples, are incomplete.

At high redshifts ($z>$2), discrepancies across different tracers become increasingly pronounced, appearing not only between UV and radio/submillimeter estimates but also among optical–IR measurements, with differences reaching up to two orders of magnitude by $z>$5, particularly in estimates of the dust-obscured SFRD \citep{Liu26}. Consequently, establishing a complete and reliable census of the cosmic star formation budget remains a major challenge for extragalactic astronomy.

Radio continuum (RC) emission from star-forming galaxies (SFGs) comprises two main components: non-thermal synchrotron radiation by cosmic ray electrons (CRE) {accelerated by supernova shocks and thermal free-free emission} from thermal electrons in ionized gas. In star-forming galaxies, both the thermal and non-thermal radiations are mainly generated in star-forming regions and hence are used as dust-unbiased tracers of young massive SFR \citep[see e.g.,][]{Condon:92,murphy:11,Taba17}. 

The radio continuum emission provides several key observational advantages: it is effectively immune to dust obscuration, largely unaffected by contributions from older stellar populations, can be measured uniformly across wide areas of the sky, and can probe cosmological distances beyond those accessible by optical/UV tracers {by detecting early dusty systems}. 

Traditionally, the tight correlation between the infrared (IR) and radio synchrotron luminosities of SFGs, known as the IR–radio correlation (IRRC) and commonly parameterized by $q_{\rm IR}$ {(i.e. the ratio of the IR to radio luminosity)}, has been used to calibrate radio-based SFRs. However, the IRRC itself remains a subject of active debate. For instance, using a stellar-mass–selected sample, \citet{Delvecchio-2021} reported that the IRRC primarily evolves with stellar mass ($M_{*}$), with more massive galaxies exhibiting systematically lower values of $q_{\rm IR}$, and also found a secondary, weaker dependence on redshift. In contrast, \citet{Cook24}, based on a radio-detected sample, found that $q_{\rm IR}$ shows little to no evolution with redshift, but exhibits a mild dependence on stellar mass. 


Moreover, most studies of the IRRC and radio-derived SFRs adopt the simplifying assumption that the radio continuum spectrum of SFGs follows a single power law. However, recent deep observations with LOFAR, uGMRT, MeerKAT, and the VLA have shown that, on average, SFGs exhibit flatter spectra at low frequencies ($\nu < 1$\,GHz) than at higher frequencies \citep{An21, Bonato21, An24}. This curvature likely arises from spectral ageing caused by the radiative energy losses of cosmic-ray electrons, which steepens the spectrum at high frequencies, and thermal free–free absorption, which flattens it at low frequencies \citep{Sweijen22, An24}. Consequently, the widespread use of a single power-law spectral model in IRRC and radio-based SFR studies introduces systematic uncertainties that must be carefully accounted for in future analyses.

 In this chapter, we emphasize the importance of broad-band radio observations of SFGs as a fundamental step to dissect their energetic thermal and non-thermal processes, calibrate the SFR independently from the IRRC, and study the cosmic SFH. We first explain possible variation in the radio spectrum of the SFGs over frequency and cosmic time, related astrophysics, and the importance of the {SKA Observatory (SKAO, \S~\ref{sec:astro})}. Then, we examine how the RC emission can be used to trace the star formation history of the Universe (\S~\ref{sec:sfhu}) and then discuss the prospect for SKA's AA4 design baseline (\S~\ref{sec:prospects}). {Throughout this chapter, we adopt the spectral-index convention $S_\nu\propto\nu^{-\alpha}$, and assume a flat $\Lambda$ cold dark matter ($\Lambda$CDM) cosmology with Hubble constant $H_{0}=67.27$\,km\,s$^{-1}$\,Mpc$^{-1}$, matter density parameter $\Omega_{\rm m}=0.32$, and cosmological constant density parameter $\Omega_{\Lambda}=0.68$ \citep{Planck2016}.}

\section{Broad-Band Radio Continuum Emission from Distant Star-forming Galaxies}
\label{sec:astro}
Studying the spectral energy distribution (SED) of galaxies provides fundamental information on the origin, energetics, and physics of
{their} electromagnetic radiation in general. The shapes of the
SEDs usually reflect the radiation laws and the underlying 
processes that affect related physical parameters. 
Thanks to coherent multi-band
observations with space telescopes like {\it IRAS}, {\it ISO}, {\it Spitzer}, and {\it Herschel}, the IR SEDs of galaxies
have been dissected and integrated as an SFR tracer. In radio, however, most of the surveys have targeted a single radio frequency/band (mainly 1.4 GHz), prohibiting a coherent radio-SED analysis for galaxy samples.
This has been mainly because of a simple assumption under which the non-thermal radio spectrum has a fixed power-law index of $\alpha_{\rm nt}\simeq$ 0.8 (for $S\propto \nu^{-\alpha_{\rm nt}}$).
However, in many cases, this assumption cannot explain the observed complexity of the radio spectra and the variation in radio spectra between SFGs
\citep[e.g.,][]{Clemens:10,Marvil,An21, Dey:22, An24, dey:24,grundy:25}.

In general, the broad-band rest-frame RC SED of galaxies can be divided into two main domains: the non-thermal domain at $\nu<$ 10 GHz and the thermal domain at frequencies $10-20<\nu<100$ GHz. The thermal free–free absorption and ionization loss can cause curvature or flattening of the non-thermal SEDs, which occurs primarily at $\nu<1$ GHz \citep[e.g.,][]{Condon:92,Marvil}. Fig.~\ref{fig:spectra} shows a theoretical radio spectrum 
 in relation to the frequency coverage of the SKA for three redshift regimes \citep[e.g., ][]{Galvin:18}. 
Toward higher frequencies $\nu>$10\,GHz, there is a possible contribution from 
{the so-called anomalous dust emission \citep[AME; not seen in Fig.~\ref{fig:spectra}, e.g.,][]{scaife:10,
Fern,Yoon01.2026.SKA}}. A robust SED survey of a sample of nearby SFGs in the mid-RC frequency range (1-10\,GHz), where the effects of absorption and AME are minimal and the emission can be assumed to be optically thin, showed that $\langle\alpha_{\rm nt}\rangle=0.97\pm0.16$ \citep{Taba17}. This indicates that, on average, the dominant cooling mechanism of CREs in nearby SFGs is synchrotron (or inverse Compton) loss. They also showed that the luminosity obtained by integrating the mid-RC SED (MRC) provides a more robust SFR tracer than the 1.4\,GHz luminosity.

One of the most pressing questions in tracing the SFR in the early universe is the possibility of the evolution of the radio SEDs. Theoretically, the shape of the SED can change with redshift depending on the evolution of the ISM properties, such as cooling/acceleration of CREs, magnetic fields, and gas density, which we explain as follows.

\subsection*{CREs cooling and acceleration}

The radio spectrum of the synchrotron emission depends on the energy distribution of the underlying relativistic electrons. As these electrons gyrate along magnetic field lines, they lose energy, and the more energetic electrons lose energy more quickly. As the energy distribution of the electrons changes, so does the synchrotron radio spectrum. A spectral break of $\Delta\alpha=-0.5$ occurs where the spectrum steepens towards higher frequency. 
Larger, and therefore more massive galaxies, can {retain} relativistic electrons with their magnetic fields longer, hence they have time to age before finally escaping the galaxy \citep[e.g.][]{Heesen2022b, An24}. If galaxies are more compact at high redshift the synchrotron spectra may not steepen as the electrons do not have time to age before escaping the galaxy. 

The relativistic electrons can also interact with photons in the local radiation field, causing the photons to up-scatter to higher energies and the electrons to lose energy, i.e., suffer inverse Compton (IC) losses. The most intense radiation field at these radio frequencies is typically the cosmic microwave background (CMB), although in theory, the ISM could cause a high photon density. The intensity of the CMB increases rapidly, as approximately $(1+z)^4$, so it was predicted that radio-loud AGN \citep{afonso:15} and SFGs \citep{murphy:2009} would both demonstrate IC losses at high redshifts. \cite{Whittam:25} found a lower radio luminosity at 1.4\,GHz for a given SFR for Lyman-break galaxies at $3<z<5$, linking it to a dominant IC scattering of the CMB. We, however, note that a lack of radio emission is found in local analogous of high-z Lyman-break galaxies as well, which is linked to  escape of cosmic-ray electrons via diffusion or galactic-scale outflows or a dominant young stellar population, which has not yet established a strong supernova-driven synchrotron continuum \citep{Greis,sebastian}.    
The spectral break from IC losses is the same as caused by spectral ageing ($\Delta\alpha=-0.5$) hence it is sometimes difficult to distinguish between these two processes. However, if higher redshift SFGs are more compact than local galaxies then they may be less likely to exhibit a spectral break due to synchrotron ageing and more likely to do so due to IC losses. The up-scattering of CMB photons to higher energies may reveal itself in non-stellar emission in the UV/X-ray SED of SFGs. 

\subsection*{Galactic magnetic fields}

The synchrotron emission from SFGs is totally dependent on the large scale galactic magnetic fields. A `typical' magnetic field\footnote{see \cite{Beck_15} for a review on magnetic fields in galaxies} strength in nearby SFGs is around $B=10\,\mu$G \citep[][]{Taba17}   
and can reach up to an order of magnitude higher in nuclear starbursts \citep{murphy:11}. A stronger magnetic field actually decreases the relativistic electron lifetime by increasing the gyro-frequency and hence increasing the rate at which electrons lose energy (i.e., via synchrotron loss) provided that the magnetic field is uniform. In turbulent fields, however, CREs scatter off the pitch angles of the field preventing efficient synchrotron cooling. This explains the flat energy index of CREs in star-forming regions where the strong magnetic field is turbulent and tangled \citep{Hassani,taba_22,nasir}. 
Some studies suggest that magnetic field strength increases with redshift \citep[e.g.][]{Bernet:08}, {{which shortens}} the lifetime of the synchrotron emission by accelerating electron aging. {{However, recent observations suggest that the increase in magnetic field strength with redshift is driven by turbulence injected by star formation activity \citep[small-scale turbulent dynamo,][]{taba_2025}, which reduces the efficiency of  synchrotron cooling of CREs at high redshift.}}  

\subsection*{Gas density and temperature}

The interstellar medium (ISM) is a key component of the baryon cycle within galaxies and influences all {tracers} of star formation in galaxies. In the radio regime it is naturally responsible for free-free emission (from the warm $T\sim 10^4\,$K ionized gas in the HII regions) as well as free-free absorption of both free-free photons and {{intermixed}} synchrotron photons at lower frequencies. 

The temperature of the HII regions affects both the total free-free luminosity normalisation {{(see the companion chapter by \citealt{Algera01.2026.SKA}})} as well as the optical depth. The latter is also {{affected}} by the electron density via the emission measure. Hence, if ISM densities and temperature increase at higher redshift then their radio spectra could flatten or turn-over at higher frequencies compared to otherwise similar local galaxies. Higher temperatures, potentially expected at higher redshifts could increase the free-free normalisation too.


Assuming a fixed SED shape (or $k$-correction) can lead to erroneous calibration of the SFR \citep[e.g.,][]{An21,taba_2025} demonstrating the importance of Deep multi-band observations. 
%
Investigating the radio SEDs at different redshifts is the first step to understanding the cosmic evolution of the $k$-correction needed to obtain the SFR using radio luminosities. 



\begin{figure}
\centering
\includegraphics[width=0.7\columnwidth]{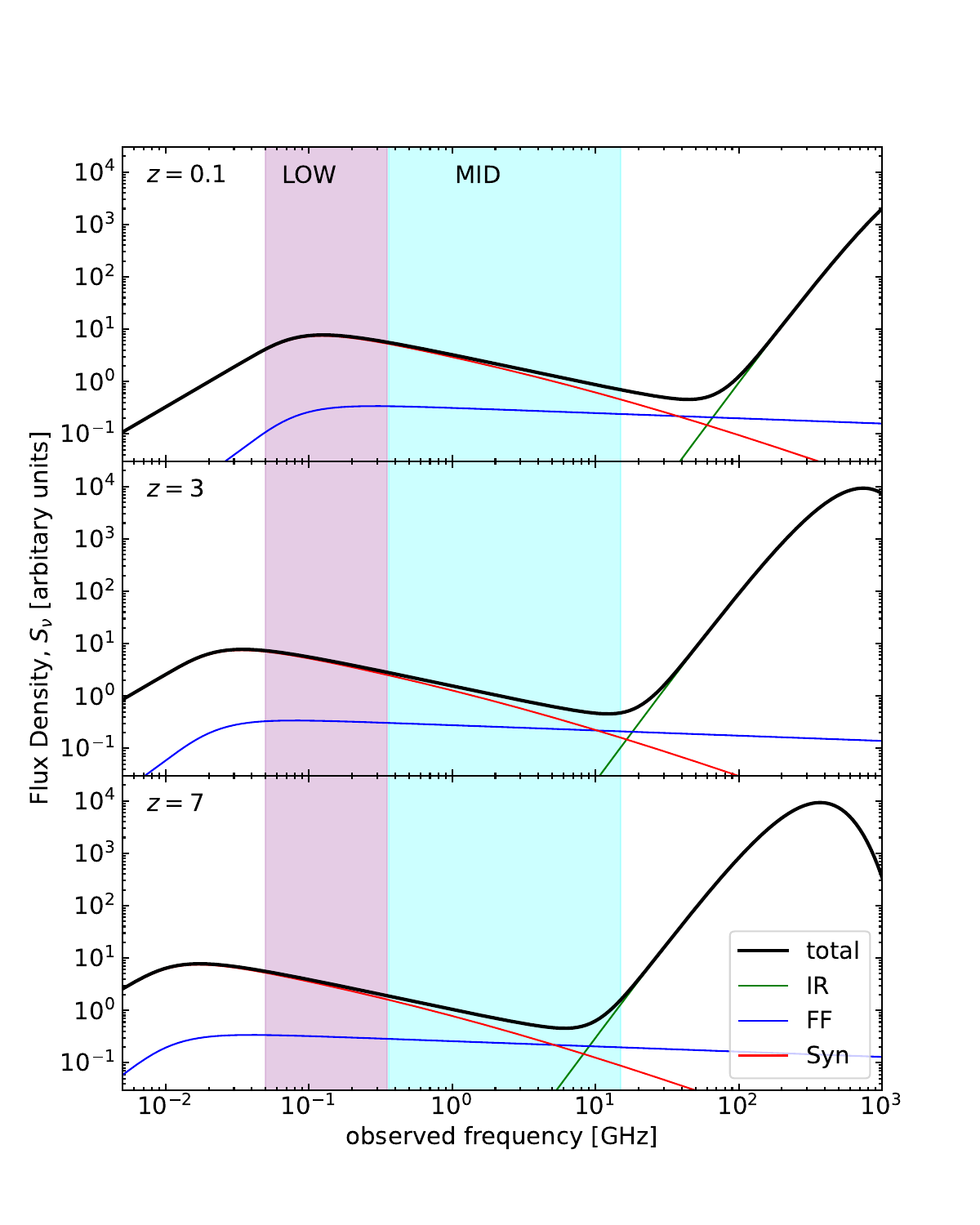}
\caption{
Theoretical radio SED of a typical star-forming galaxy in the observed-frame at different cosmic times ($z=0.1$ or local, $z=3$ or cosmic noon, and $z=7$ at the epoch of reionisation) assuming no evolution. Also shown are the SKA-LOW (pink) and SKA-MID (cyan) frequency coverage. The radio spectra comprise a free-free (FF) and synchrotron (Syn) components, both of which are suffering the same free-free absorption at $\nu<1$\,GHz, as well as a thermal grey-body IR component (IR) from cold dust. The synchrotron emission steepens around $15\,$GHz rest-frame due to aging and/or inverse Compton losses. The flux density units are arbitrary. 
}
\label{fig:spectra}
\end{figure}

\subsection{Observed evolution of the radio spectra and the role of the SKAO}

\begin{figure*}
\centering
\resizebox{\hsize}{!}{\includegraphics[width=0.49\columnwidth]{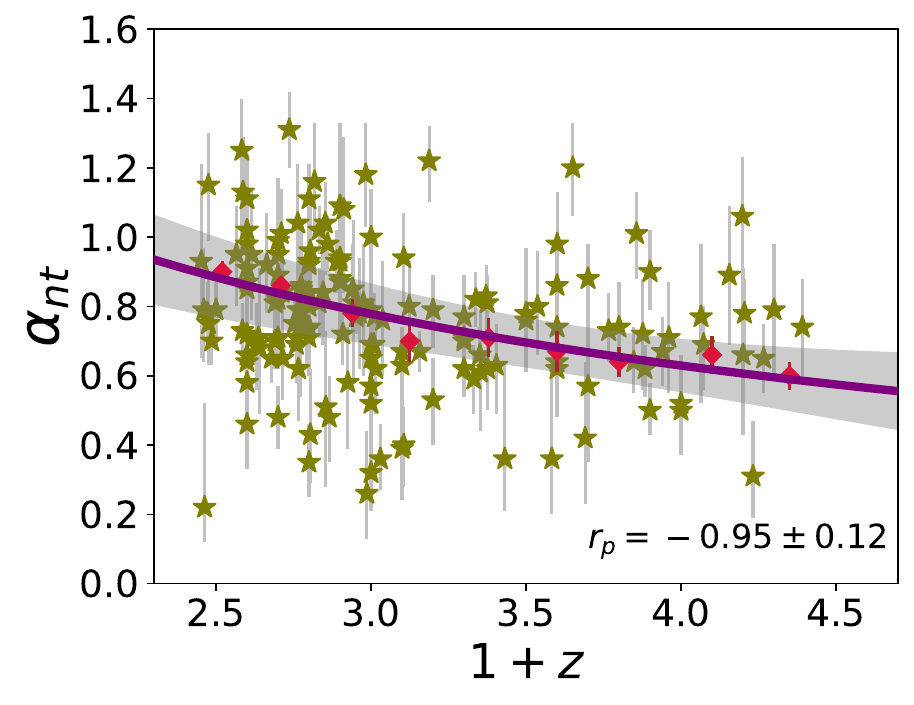}{\includegraphics[width=0.51\columnwidth]{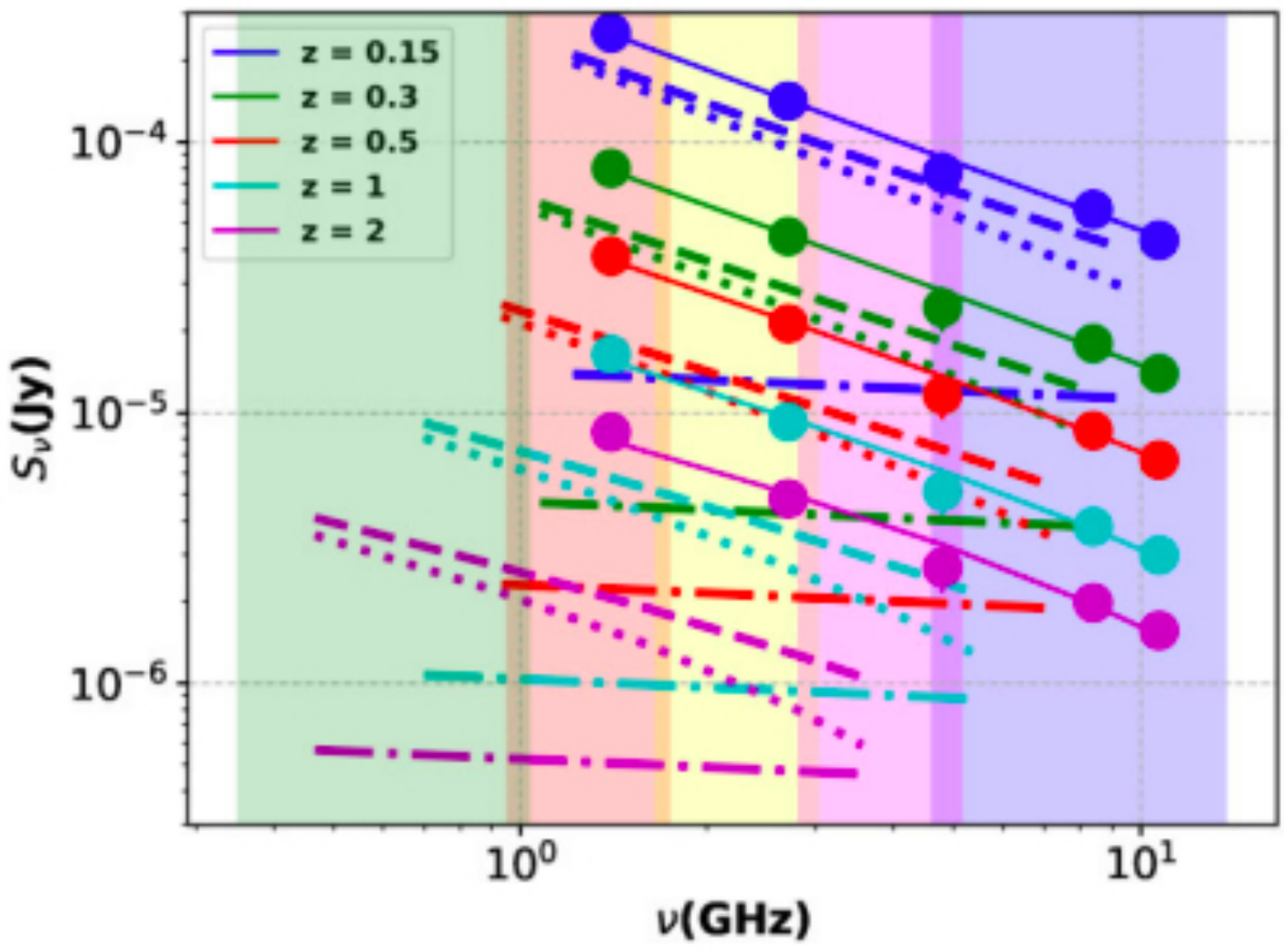}}}
\caption{{\it Left-}
non-thermal spectral index $\alpha_{\rm nt}$ against redshift z for MIGHTEE-COMOS SFGs located at 1.5 < z < 3.5 \citep{taba_2025}. The curve show best fit to the mean values in redshift bins of 0.2 (red points) with Pearson correlation coefficient, $r_p$. Gray-shaded area show the 99\% confidence bound. {\it Right-} The fitted rest-frame SEDs (solid lines) of a M51-like galaxy at
five different redshifts adapted from \cite{ghasemi}. Points show the rest-frame integrated flux densities at
1.4, 2.7, 4.8, 8.4, and 10.7 GHz at different
redshifts. Also shown
are the corresponding SEDs in the observer-frame (dashed lines) and their
thermal (dashed-dotted curves) and non-thermal (dotted curves) components.
Vertical coloured shades show the SKA frequency bands 1 (green), 2 (red), 3
(yellow), 4 (pink), and 5 (blue).}
\label{fig:SED}
\end{figure*}

Unlike the assumption used in Fig.~\ref{fig:spectra}, observations with SKAO pathfinders and precursors such as VLA and MeerKAT indicate variations in the radio spectral index of SFGs measured at high-z. {Studying radio-selected M51-like galaxies (SFR\,$\sim$\,2-10\,M$_\odot$yr$^{-1}$) in the GOODS-N field, \cite{murphy_17} found a flatter spectral index ($\alpha_{\rm 1.4GHz}^{\rm 10GHz}\simeq$0.61) at $\langle z \rangle =1.24 \pm 0.15$ than that of nearby galaxies \citep[$\alpha_{\rm nt}\simeq$1,][]{Taba17}.  
A more intense} flattening was reported in highly star-forming galaxies (SFR\,$\geq$ 100\,M$_\odot$yr$^{-1}$) at high redshifts,  in the VLA-COSMOS 3\,GHz Large Project by \cite{tisanc}. These authors found a mean spectral index of $0.42$ ({below a rest-frame frequency of 4.3\,GHz}) at $\langle z\rangle =1.7 \pm 0.6$, which is {flatter than that of galaxies with lower SFR presented by \cite{murphy_17} at about similar redshifts}. More recently, a detailed SED analysis of the MIGHTEE-COSMOS SFGs finds a systematic flattening of $\alpha_{\rm nt}$  with redshift {\citep[see Fig.\ref{fig:SED}-left]{taba_2025}}. The cosmic evolution of the SFR and the fact that the synchrotron spectrum becomes flatter with an increasing specific SFR \citep{taba_2025} or $\Sigma_{\rm SFR}$ \citep[][]{Taba17}, is most likely causing this SED variation with $z$. This agrees with the scenario in which the number of high-energy CREs increases at higher $z$ because of the enhanced star formation activity and its resulting CREs acceleration processes such as Fermi-II acceleration in moving turbulent magnetic fields. 

In an attempt to assess the capability of the SKAO in tracing back the physical processes in early galaxies,  \cite{ghasemi} simulated the thermal and non-thermal RC emission of M51 analogs at high-z. Taking into account the evolution of the SFR and using the correlation between $\alpha_{\rm nt}$ and $\Sigma_{\rm SFR}$, they found the same spectral index as that reported by \cite{murphy_17} for similar population of SFGs observed in the GOODS-N field. For integrated studies, \cite{ghasemi} provide the following expression to trace back the radio flux of a nearby SFG {to} its earlier cosmological times: 

\begin{equation}
    S(z)=S(0)\,\frac{\rm SFR(z)}{\rm SFR(0)}\,\frac{D^2(0)}{D_L^2(z)}\, (1+z)^{(1-\alpha)},
\end{equation}
with $S(z)$ the observed integrated flux density of the galaxy at redshift $z$. In the above relation, $S(0)$, SFR(0), and D(0) are the integrated flux density,
SFR and distance of that galaxy (at z$\sim$0). Using multi-band RC flux densities, Eq(1) allows construction of SEDs of SFGs at earlier cosmic times or redshifts. In { Fig.~\ref{fig:SED}-right}, we show, for instance, the evolution of the mid-RC SED of an M51 analog out to $z=2$ and the different bands of the SKAO covering it.  {At} higher redshifts ($z>2$), the SKA-LOW will be needed to cover the rest-frame mid-RC SEDs relevant to our MRC calibration of the SFR/SFH (see below). 


\section{Radio luminosity to Star Formation Rate Calibration}
\label{sec:calibration}

\begin{figure}
\centering
\includegraphics[width=0.5\columnwidth]{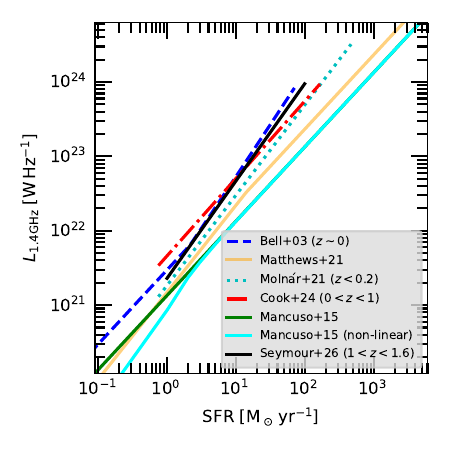}
\caption{
Selection of the commonly used 1.4\,GHz to star formation rate conversion factors \citep{bell:03,Mancuso:15,Matthews21,Molnar-2021,Cook24,seymour:26}. If determined over a limited redshift range, this is indicated in the legend. If determined over a limited range of SFRs or radio luminosities, this is indicated by the truncated lines. 
}
\label{fig:lradsfr}
\end{figure}

{Measuring the rate at which massive stars form in galaxies is key to understanding the formation and evolution of galaxies.
Various emission lines and continuum photometry have been used so far as SFR diagnostics, each with advantages and shortcomings \citep{ken}. The most frequently used tracers, H$\alpha$ and UV (rest frame $125-250\,$nm) emission, are directly related to the massive star-formation process, but they
could be obscured or attenuated by interstellar dust. This has
motivated the use of hybrid star-formation tracers combining two
or more different tracers, including IR emission, to correct for
the dust attenuation. The use of IR emission itself as an SFR
tracer is affected by a contribution from other sources/
mechanisms unrelated to massive star formation, such as
interstellar dust heating by solar-mass stars \citep{Calzetti}
 and emission from the atmosphere of carbon
stars \citep[mainly in mid-IR, e.g.,][]{verley,taba_10}. 
As mentioned earlier, the RC emission is not attenuated by dust and no other tracer is needed to be combined with. Hence, the radio SFRs have the potential to provide a more precise measure of the rate of massive star formation
in a galaxy than the common non-radio SFRs. Similar to the H$\alpha$ emission, the thermal RC component traces the ionized gas surrounding young massive stars (HII regions) with lifetimes of $\sim$\,3-10\,Myr. Evolving fast to the supernova phase (in a few million years), massive stars are then traced by the non-thermal RC component. However, as extragalactic star-forming regions are often observed as complexes of the HII and supernova phases, they radiate a mix of both thermal and non-thermal RC components \citep{taba_07,taba_13}. In star forming regions with typical field strengths of  B$>20\,\mu$G, the non-thermal emission at $\nu=1.4$\,GHz also traces very recent SFR $\leq\,10$\,Myr (the synchrotron lifetime is given by $t_{\rm syn}{\rm \sim 1.06\times 10^9 yr \, (\frac{B}{\mu G})^{-1.5}\, (\frac{\nu}{GHz})^{-0.5}}$). Far away from star forming regions, the synchrotron of diffused CREs can still trace the SFR back to longer time scales. However, we note that the SFR--synchrotron correlation becomes super-linear in resolved studies due to diffusion of CREs \citep{heesen14,taba_22}. On the other hand, in unresolved galaxy-scale studies, this correlation (SFR vs non-thermal RC) is often found to be sub-linear \citep[e.g.,][]{chi,Molnar-2021,Basu15}, meaning that galaxies with higher SFR produce synchrotron emission in excess of a linear scale. {Considering} that the galaxies with higher SFR also have flatter non-thermal spectral index, \cite{Taba17} linked their excess synchrotron emission to a non-thermal feedback: In addition to injecting CREs, massive star formation amplifies turbulent magnetic fields helping the onset of wind and outflow structures in SFGs. 



As mentioned in Section~$\S$\ref{sec:introduction}, 
calibrating the SFR using the IRRC is complicated by its observed change with redshift \citep[e.g.,][]{Delhaize-2017, Magnelli-2015} which can depend on galaxy mass and type \citep[e.g.,][]{Delvecchio-2017, Delvecchio-2021, Molnar-2021, DeZotti-2024}. We present a range of the current radio-SFR conversions in Fig.~\ref{fig:lradsfr}, demonstrating the scatter. Taking into account the complexity of the radio--IR correlation, \cite{Taba17} presented a more direct use of the radio continuum emission as an SFR tracer in galaxies based on a detailed study of their radio SEDs. They showed that the scatter in the SFR calibration is reduced by using their integrated mid-radio (1-10\,GHz) luminosity (MRC) instead of the monochromatic luminosities at 1.4 or 4.8\,GHz in nearby galaxies.
%
%
%

%
\subsection{SFR Calibrations at High Redshifts}\label{sec:SFR calibarion(z)}
Calibrating SFRs in the early universe is pivotal for reconstructing the cosmic star formation history and understanding galaxy evolution during peak epochs of activity. At high redshifts, galaxies exhibit elevated star formation, often obscured by dust and shaped by evolving interstellar medium (ISM) conditions. {Deep radio continuum observations offer a dust-unbiased window into the early universe.} 
Using combined observations from MeerKAT, VLA, and GMRT, and a Bayesian method, \cite{taba_2025} modelled the rest-frame radio SEDs for 160 star-forming galaxies in the COSMOS field, spanning the redshift range $1.5 < z < 3.5$. This enabled the decomposition of the radio continuum into synchrotron and free–free emission components. {Two important outcomes of this study are 1) radio SEDs of most of the SFGs do not need a thermal component  and 2) the SEDs} evolve with redshift due to the evolution of SFR. Taking into account the resulting variation in the $k$-correction, this study finds that the radio--IR correlation remains statistically invariant across the targeted redshift range.  The SFR calibration with the MRC results in a relation that is closer to being a linear one, and a scatter smaller than that calibrated with monochromatic radio bands {(see Fig.~\ref{SFRTIRMRC})}:  

\begin{equation}\label{eq:1}
\left(\dfrac{\rm SFR_{TIR}}{\rm M_{\odot} \; yr^{-1}} \right)  \; =  \; 10 ^{\; (0.29 \; \pm \; 0.13)} \; \times \;
\left(\dfrac{\rm SFR_{1.3} }{\rm M_{\odot} \; yr^{-1} }\right)^{ (0.75 \; \pm \; 0.04)}, 
\end{equation}

\begin{equation}\label{CallibMRC}
 \left( \dfrac{\rm SFR_{TIR}}{\rm M_{\odot} \; yr^{-1}} \right) = 10^{(0.38\pm0.10)} \;   \left( \dfrac{\rm SFR_{MRC}}{\rm M_{\odot} \; yr^{-1}} \right)^{(1.00 \; \pm \; 0.04)}.
\end{equation}
{This} is expected because a more complete sampling of CREs with different energies is attained using the integrated radio SED than a single radio band.


\begin{figure*}[ht!]
	\begin{center}
		\resizebox{\hsize}{!}{\includegraphics[width=6cm]{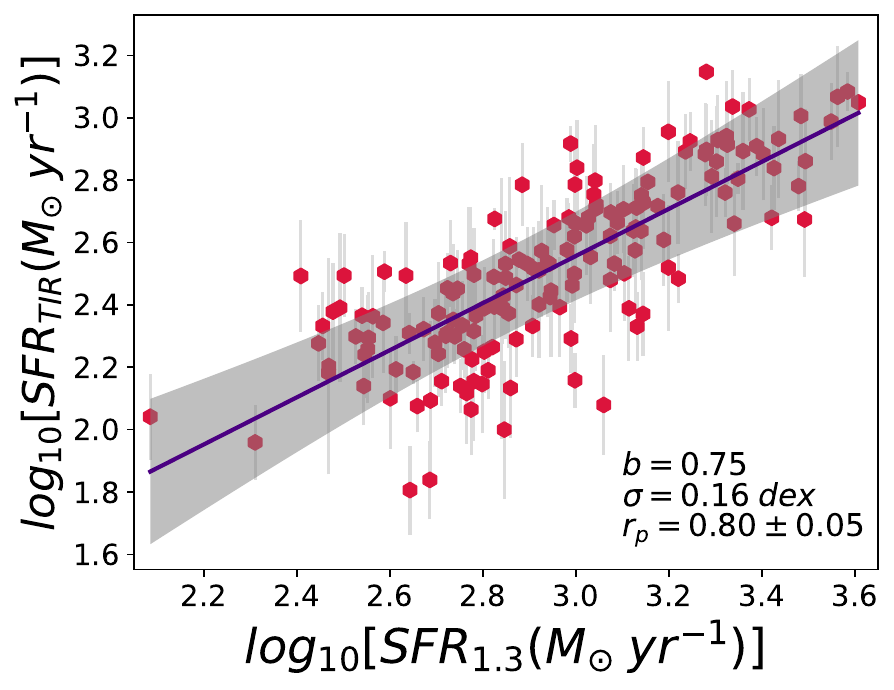}
        \includegraphics[width=6cm]{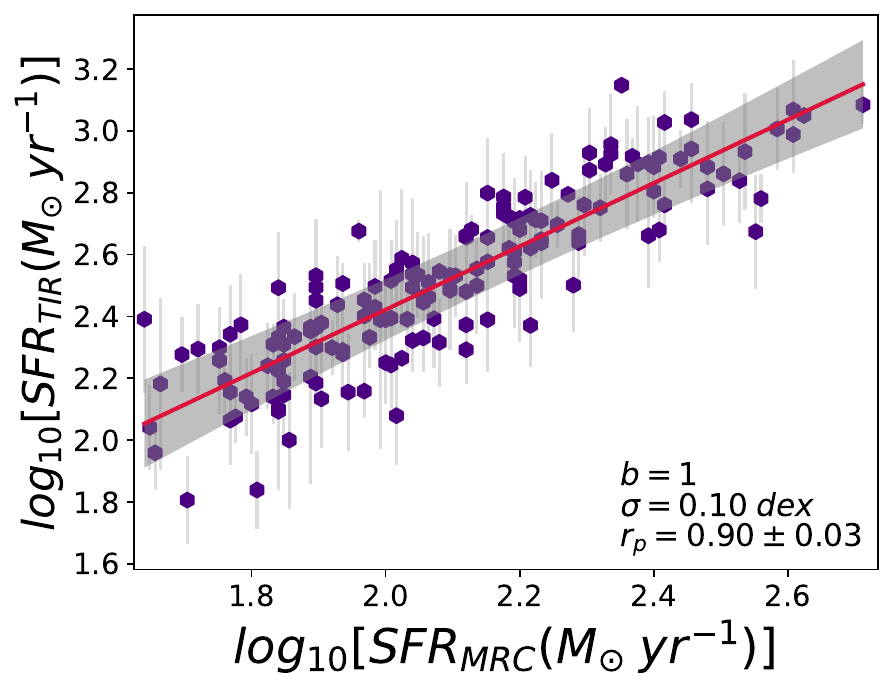}} %
				\caption[]{The SFR of the MIGHTEE-COSMOS SFGs at $1.5 < z < 3.5$ measured using the TIR luminosity against the SFRs calibrated in radio adapted from \cite{taba_2025}: {\it left-} 1.3\,GHz  and {\it right-} MRC. Solid lines show the linear regression fits to the data in all panels. The gray shaded areas show their 99\% confidence bounds.} 
		\label{SFRTIRMRC}
	\end{center}
\end{figure*}


\section{From Star Formation Rate to Cosmic Star Formation History}
\label{sec:sfhu}

The measured SFR can be used to infer the cosmic SFH if the radio luminosity function and its evolution are known. In what follows, we first review updates on the radio luminosity function studies and then discuss the expected cosmic SFH. 

\subsection{Radio Luminosity Function and its evolution}\label{sec:RLF}
{The radio luminosity function (RLF) of star-forming galaxies (SFGs), $\Phi(L,z)$, provides a fundamental statistical description of their space density and underpins radio-based measurements of the cosmic star formation history. In the past decade, major progress has been driven by deep, wide-area, and multi-frequency radio surveys, enabling a more physically consistent view of the RLF across both luminosity and redshift.

At low redshift, recent studies combining LOFAR, GMRT, and VLA data have established a consistent local RLF across frequencies from $\sim150\,\mathrm{MHz}$ to $1.4\,\mathrm{GHz}$ \citep[e.g.,][]{2016MNRAS.462.1910H,2019A&A...622A..17S,2016MNRAS.457..730P}. These results show that the RLF is well described by a modified Schechter function \citep{1990MNRAS.242..318S}:
\begin{equation}
\label{eq:LRLF}
\Phi_{0}(L)\,d\log_{10}L
=
\Phi_{\star}
\left(\frac{L}{L_{\star}}\right)^{1-\beta}
\exp \left[
-\frac{1}{2\sigma^{2}}
\log_{10}^{2}\left(1+\frac{L}{L_{\star}}\right)
\right]
d\log_{10}L ,
\end{equation}
where $L_{\star}$ characterizes the knee of the RLF, and $\beta$, $\sigma$, and $\Phi_{\star}$ describe its shape and normalization. 

{As discussed in Sections~$\S$\ref{sec:astro} and  $\S$\ref{sec:calibration}, a  key advance enabled by multi-frequency data is the direct constraint of radio SEDs}, reducing reliance on simple power-law assumptions and improving $k$-corrections and luminosity estimates. This is particularly important for connecting low-frequency (e.g., LOFAR) and GHz-frequency (e.g., VLA, MeerKAT) measurements within a unified framework.

At higher redshifts, surveys such as VLA-COSMOS, {LOFAR Two Metre Sky Survey (LoTSS)}, and MeerKAT MIGHTEE have enabled the construction of large SFG samples with robust classifications based on multi-wavelength diagnostics \citep[e.g.,][]{Novak17,2020MNRAS.491.5911O,Cochrane23,Hale25}. The combination of multi-frequency radio data and ancillary optical/IR information allows improved separation of SFGs and AGN, which is critical for reliable RLF measurements in the sub-mJy regime.

Methodologically, there has been significant progress beyond the classical $1/V_{\max}$ estimator \citep{1968ApJ...151..393S}. Recent studies increasingly adopt kernel density estimation (KDE) and maximum-likelihood approaches with explicit completeness corrections \citep[e.g.,][]{2020ApJS..248....1Y,2022ApJS..260...10Y,Wang24,Wang26}, enabling continuous reconstruction of $\phi(L,z)$ and reducing binning-related biases.

A consistent result from recent work is that pure luminosity evolution (PLE) is insufficient to describe the observed RLF evolution. Instead, both luminosity and density evolution are required, commonly parameterized as:
\begin{eqnarray}
\label{eq:RLF}
\phi(L,z) = g(z)\,\Phi_{0}\!\left[\frac{L}{f(z)}\right]
\end{eqnarray}
where $f(z)$ and $g(z)$ describe luminosity and density evolution, respectively. This luminosity and density evolution (LADE) framework is strongly supported by recent analyses of both GHz and low-frequency datasets \citep[e.g.,][]{vanderVlugt22,Cochrane23,Wang24,Wang26}.

Despite these advances, significant challenges remain. Current surveys are still limited in probing the faint end and the knee of the RLF at high redshift, leading to degeneracies between luminosity and density evolution. Future surveys with {SKAO}, combining unprecedented sensitivity and broad frequency coverage, will be essential to fully constrain the shape and evolution of the RLF across cosmic time.}

\subsection{RLF to cosmic SFH}
With well-constrained RLFs at different redshift, the cosmic SFRD, $\psi(z)$, at a given epoch, $z$, can be estimated by integrating the RLF:
\begin{eqnarray}
\psi(z)=C\times \int_{L_{\rm min}}^{L_{\rm max}} \phi(L,z)\,{\rm SFR}(L)\,{\rm d\,log_{10}}L,
\end{eqnarray}
where $C$ is the correction factor that accounts for survey completeness and the conversion between different assumptions of the initial mass function (IMF). The term $\phi(L, z)$ represents the RLF, which can be expressed as Equation~\ref{eq:RLF} when incorporating luminosity and density evolution, as discussed in Section~$\S$\ref{sec:RLF}. The function ${\rm SFR}(L)$ provides the calibration between radio luminosity ($L_{\rm radio}$) and SFR, as described in Section~$\S$\ref{sec:SFR calibarion(z)} (see Equations~\ref{CallibMRC} and Fig.~\ref{fig:lradsfr}). In practice, this may vary with {stellar mass} and redshift. Ideally, the integration would extend over the entire luminosity range, i.e. $L_{\rm min}=0$ and $L_{\rm max}\to+\infty$. In practice, however, for flux-limited surveys, $L_{\rm min}$ and $L_{\rm max}$ correspond to the lowest and highest value of the observed RLF. If the knee of the luminosity function ($L_*$ from equation~\ref{eq:LRLF}) falls between $L_{\rm min}$ and $L_{\rm max}$ then the vast majority of the star formation is directly observed and $C$ is close to unity. 

\begin{figure}
\centering
\includegraphics[width=0.7\columnwidth]{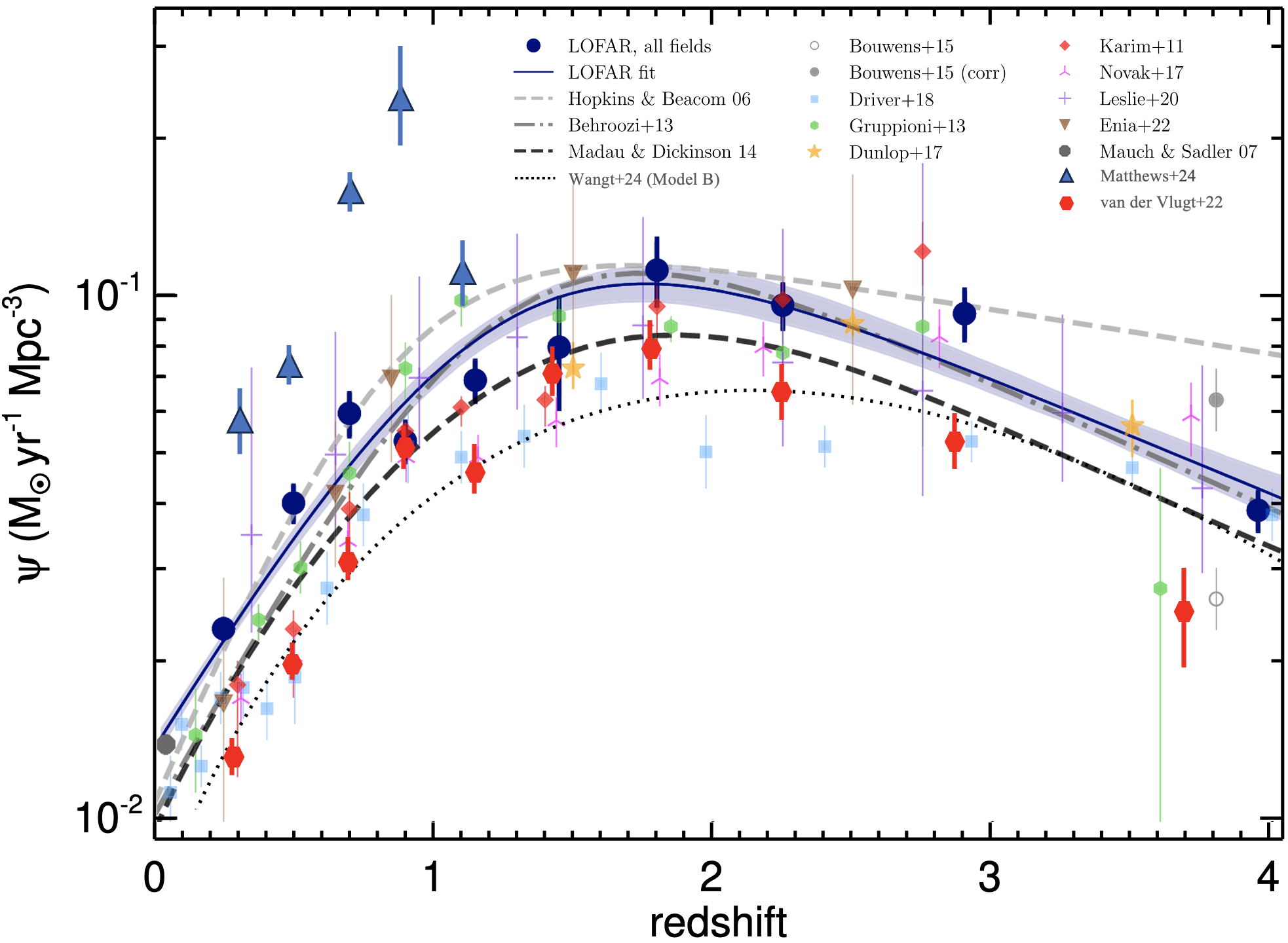}
\caption{Cosmic star formation rate density (SFRD) as a function of redshift. The figure is reproduced from \cite{Cochrane23}, who used the high-sensitivity 150\,MHz data from LoTSS to estimate the cosmic SFRD up to $z\lesssim4$. Other recent radio continuum–based measurements are overlaid and listed in the rightmost column of the legend. Except for the results from \citet[][VLA-COSMOS 1.4\,GHz]{Karim11}, \cite{Novak17} and \citet[][VLA-COSMOS 3\,GHz survey]{Leslie-2020}, and \citet[][VLA 1.4\,GHz, GOODS-N field]{Enia22}, which are already included in \cite{Cochrane23}, we additionally show the measurements from the MeerKAT DEEP2 field \citep{Matthews24} and the combined COSMOS-XS$+$VLA-COSMOS 3\,GHz data sets \citep{vanderVlugt22}. For comparison, the cosmic SFRD derived from other wavelength regimes is also shown: UV \cite[grey squares;][]{Bouwens15}, optical/NIR \cite[blue squares;][]{Driver18}, and FIR \cite[green hexagons and orange stars;][]{Gruppioni13,Dunlop17}. Previously derived empirical fits to the cosmic SFRD are overplotted from \cite{Hopkins06}, \cite{Madau14}, \cite{Behroozi13}, and \cite{Wang24}.}
\label{fig:SFH_radio_Cochrane23}
\end{figure}

Figure~\ref{fig:SFH_radio_Cochrane23} presents recent measurements of the cosmic SFRD derived from radio continuum surveys. {In addition to the discrepancies between radio-based} and UV–IR–based estimates of the cosmic SFH, significant variations are also evident among the radio-derived results themselves. For instance, at $z\lesssim4$, \citet{Cochrane23}, using high-sensitivity 150\,MHz data from the LoTSS, and \citet{Leslie-2020}, based on the VLA-COSMOS 3\,GHz survey, both report SFRD values exceeding the empirical UV–IR fit from \citet{Madau14}. In contrast, \citet{Novak17} and \citet{vanderVlugt22}, who also use or include the VLA-COSMOS 3\,GHz data, find SFRDs that lie below the same empirical relation. These inconsistencies arise primarily from differences in the adopted radio–SFR calibrations, assumptions in the construction of the RLF, as well as the observational frequencies used \citep{Leslie-2020,vanderVlugt22,Cochrane23,Matthews24}. 

This underscores the critical importance of establishing robust radio–SFR calibrations (Section~$\S$\ref{sec:calibration}) and well-constrained RLFs and their cosmic evolution (Section~$\S$\ref{sec:RLF}) before drawing firm conclusions about the cosmic SFH from the radio continuum surveys. Once surveys achieve sufficient depth to probe the full RLF shape, from the exponential cut-off through the knee to the faint-end power-law slope, the need to adopt oversimplified evolutionary assumptions such as pure luminosity or pure density evolution will be greatly reduced. {{Studying the radio SEDs is fundamentally important as it provides both robust SFR calibrations and an insight about possible variations in the k-correction needed to construct RLFs.}
In this context, the prospects offered by SKA {AA4} for tracing the cosmic SFH from the radio perspective are discussed in Section~$\S$\ref{Sec:SKA_AA4}.}

\section{Prospects with SKA AA4}\label{Sec:SKA_AA4}
\label{sec:prospects}
\subsection{Broad Frequency Capabilities of SKA AA4 in Tracing the Radio Spectra of SFGs}\label{sec:BFC_SKA}
\subsubsection{Broad-band Radio Spectral Diagnostics with SKA AA4}
Figure~\ref{fig:spectra} illustrates the broad frequency coverage of SKA-Low (50–350\,MHz) and SKA-Mid (0.35–15\,GHz), highlighting the crucial role this range plays in studying the radio spectra of SFGs, particularly at high redshift. Specifically, SKA-Low will be essential for: (1) characterising spectral curvature at $z\lesssim1.5$, as shown in the top panel of Figure~\ref{fig:spectra}; and (2) probing the mid-radio continuum (MRC) SEDs of main-sequence SFGs at $z\gtrsim3.5$. 

SKA-Mid, on the other hand, will be critical for constraining the MRC SEDs at $z\lesssim3$ and for probing the high-frequency radio regime, particularly with Band\,5a (4.6–8.5\,GHz) and Band\,5b (8.3–15.4\,GHz). These frequency bands are particularly important for: (1) investigating the radio spectra of starburst and dwarf SFGs; (2) distinguishing between radio emission–dominated and infrared (dust emission)–dominated SEDs at $z>6$; and (3) disentangling the relative contributions of free–free and synchrotron emission in SFGs. The latter is especially significant at high redshift, where free–free emission may dominate the observed flux in Band\,5a\&b, necessitating careful consideration of radio–SFR calibrations. A detailed discussion of tracing the cosmic star formation history using free-free emission is presented in \cite{Algera01.2026.SKA}.

\subsubsection{Constructing the Realistic SKA SFG Mock Samples}
{To assess the impact of both frequency coverage and survey depth on future SKA constraints on the cosmic SFH, we adopt the extragalactic continuum survey framework proposed by \citet{Prandoni01.2026.SKA}}. In this chapter, we consider both the Deep multi-band Tier and the Ultra-deep Tier survey strategies.


\begin{figure}
    \centering
    \includegraphics[width=0.99\linewidth]{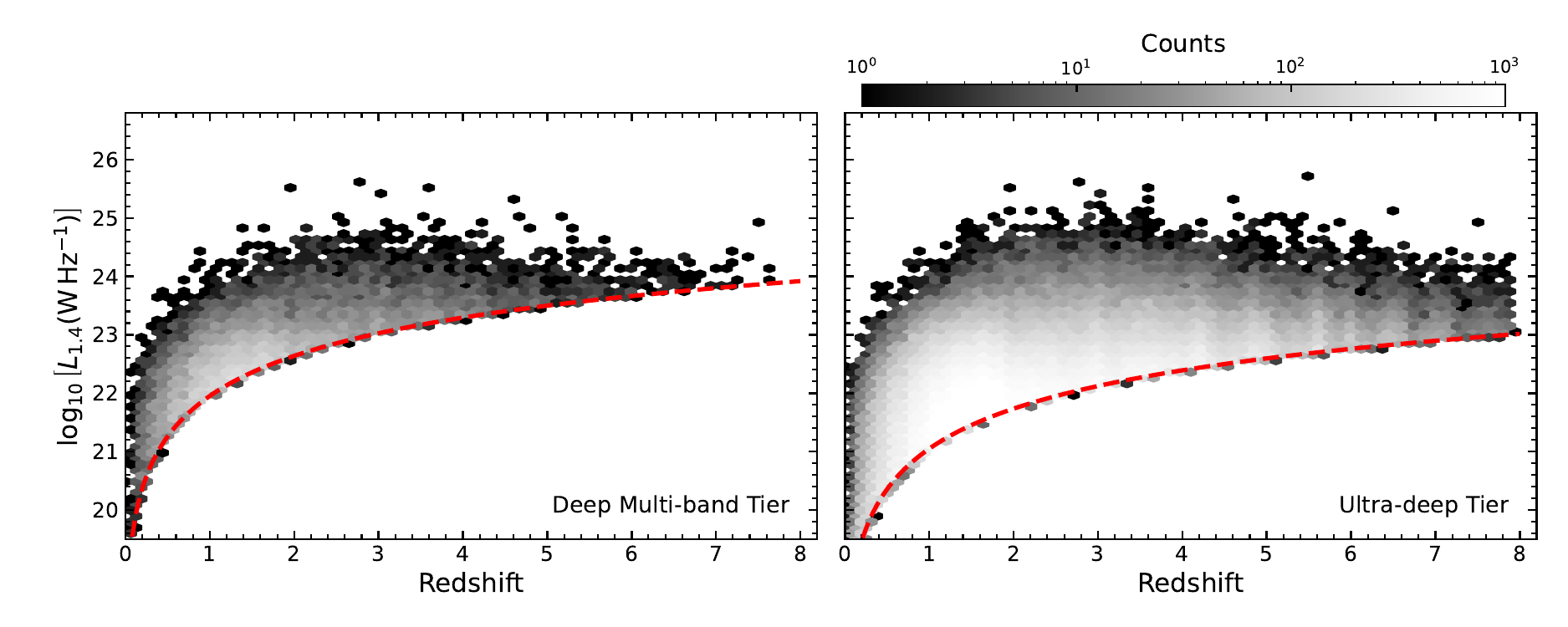}
    \caption{Rest-frame 1.4\,GHz luminosity as a function of redshift for the mock SFG sample, assuming a fixed radio spectral index of $\alpha=0.70$. {The left and right panels show the predicted source distributions for the Deep multi-band Tier and Ultra-deep Tier survey strategies, respectively, following the survey parameters presented in Table~2 and Figure~6 of \cite{Prandoni01.2026.SKA}}. The { red} dashed curves indicate the corresponding $5\sigma$ detection limits at SKA-Mid Band\,2, adopting flux-density thresholds of $S_{1.4\,{\rm GHz}}=2\,\mu$Jy for the Deep multi-band Tier and $0.25\,\mu$Jy for the Ultra-deep Tier. The colour scale represents the number of sources in each hexagonal bin.}
    \label{fig:Mock_SFG}
\end{figure}

The Deep multi-band Tier, summarised in Table~2 and illustrated in Figure~6 of \citet{Prandoni01.2026.SKA}, is designed to provide approximately beam-matched} multi-band imaging with SKA-Low and SKA-Mid, {enabling robust measurements of radio continuum SEDs for} both SFG and AGN populations. {Depending on the observing band, the survey covers an area of up to 0.25\,deg$^{2}$ and reaches an expected $1\sigma$ sensitivity of $\sim0.4\,\mu$Jy\,beam$^{-1}$ at SKA-Mid Band\,2 ($\sim1.4$\,GHz). The corresponding} depths at other frequencies are estimated by scaling from this Band\,2 sensitivity assuming a representative synchrotron spectral index of $\alpha=0.70$.

{To explore the ultimate sensitivity limits achievable with SKA AA4, we additionally consider the Ultra-deep Tier, which is planned to cover 1\,deg$^{2}$ down to a $1\sigma$ depth of $\sim0.05\,\mu$Jy\,beam$^{-1}$ at SKA-Mid Band~2. 

Mock galaxy samples for both survey tiers are constructed from the T-RECS \texttt{sfgsdeep} catalogue \citep{Bonaldi19}, which} provide mock extragalactic radio continuum skies over 150\,MHz--20\,GHz, thereby covering most of the SKA-Low frequency range and the full SKA-Mid range considered here. The simulations include radio continuum emission from both SFGs and AGN, and are designed to reproduce the observed evolution of radio source populations across a broad range of frequencies and redshifts. {Following \citet{Bonaldi19}, we treat all sources in the \texttt{sfgsdeep} catalogue as SFGs, although it also includes a contribution from radio-quiet (RQ) AGN. The information available in the mock catalogue is insufficient to separate RQ AGN from SFGs, as such a classification would require additional multi-wavelength diagnostics. We note that this RQ AGN contribution may affect the predicted source number densities and the derived RLFs. In addition, the tabulated flux densities in the mock catalogue do not include measurement uncertainties. Therefore, the mock-catalogue-based forecasts presented in this section should be interpreted as illustrative predictions of the scientific potential of SKA AA4, aimed at assessing the impact of broad frequency coverage and survey sensitivity, rather than as precise predictions of the final SKA AA4 constraints.

To construct realistic SKA mock samples, we first select all sources within a circular region corresponding to the survey area and then apply a $5\sigma$ flux-density threshold at 1.4\,GHz, which is adopted as the reference selection band. For the Deep multi-band Tier,} adopting a flux-density limit of $S_{1.4\,\rm GHz}\ge2\,\mu$Jy yields a sample of 17,019 sources, including 2,016 at $z>3$ and 237 at $z>5$. Given the matched-depth multi-frequency survey framework adopted here, the majority of these sources are expected to be detected at $>5\sigma$ across the SKA-Low and SKA-Mid bands. 

{For the Ultra-deep Tier, adopting a flux-density limit of $S_{1.4\,\rm GHz}\ge0.25\,\mu$Jy yields a sample of 258,878 sources, including 57,052 at $z>3$ and 10,248 at $z>5$. Unlike the Deep multi-band Tier, the Ultra-deep Tier is used here as a sensitivity-driven reference case, since matched-sensitivity coverage across the full SKA frequency range is not achievable owing to source-confusion limits at low frequencies.

Figure~\ref{fig:Mock_SFG} presents the distributions of rest-frame 1.4\,GHz luminosity as a function of redshift for the Deep multi-band and Ultra-deep mock SFG samples, where the luminosities are estimated assuming a fixed radio spectral index of $\alpha=0.70$. These two samples are used in the following sections to forecast how SKA AA4 can constrain the RLF and, subsequently, the cosmic SFH under different survey configurations. We emphasise that these forecasts are intended as illustrative predictions of the scientific potential of SKA AA4.}

\subsubsection{The Role of SKA Broad Frequency Coverage in Rest-Frame Radio Luminosity Estimates}
To quantify {how SKA AA4's broad frequency coverage improves rest-frame radio luminosity estimates, and hence radio-based constraints on the cosmic SFH, we analyse the Deep multi-band mock SFG sample constructed in the previous subsection (left panel of Figure~\ref{fig:Mock_SFG}).}

We then compute the rest-frame 1.4\,GHz luminosity for each SFG, applying the $k$-correction under three assumptions about the radio spectral index: (1) a fixed spectral index of $\alpha=0.70$, representative of the case where only SKA-Mid Band\,2 observations are available; (2) a two-point spectral index measured {from the T-RECS tabulated flux densities at} 780\,MHz and 1.4\,GHz, corresponding to the case where both SKA-Mid Bands\,1 and 2 are available; and (3) a best-fitting spectral index obtained through Bayesian fitting of the radio SED using the {T-RECS tabulated flux densities across the full frequency range sampled by the catalogue, from 150\,MHz to 20\,GHz, which covers most of the SKA-Low frequency range and the full SKA-Mid range considered here.}

\begin{figure}
    \centering
    \includegraphics[width=0.49\linewidth]{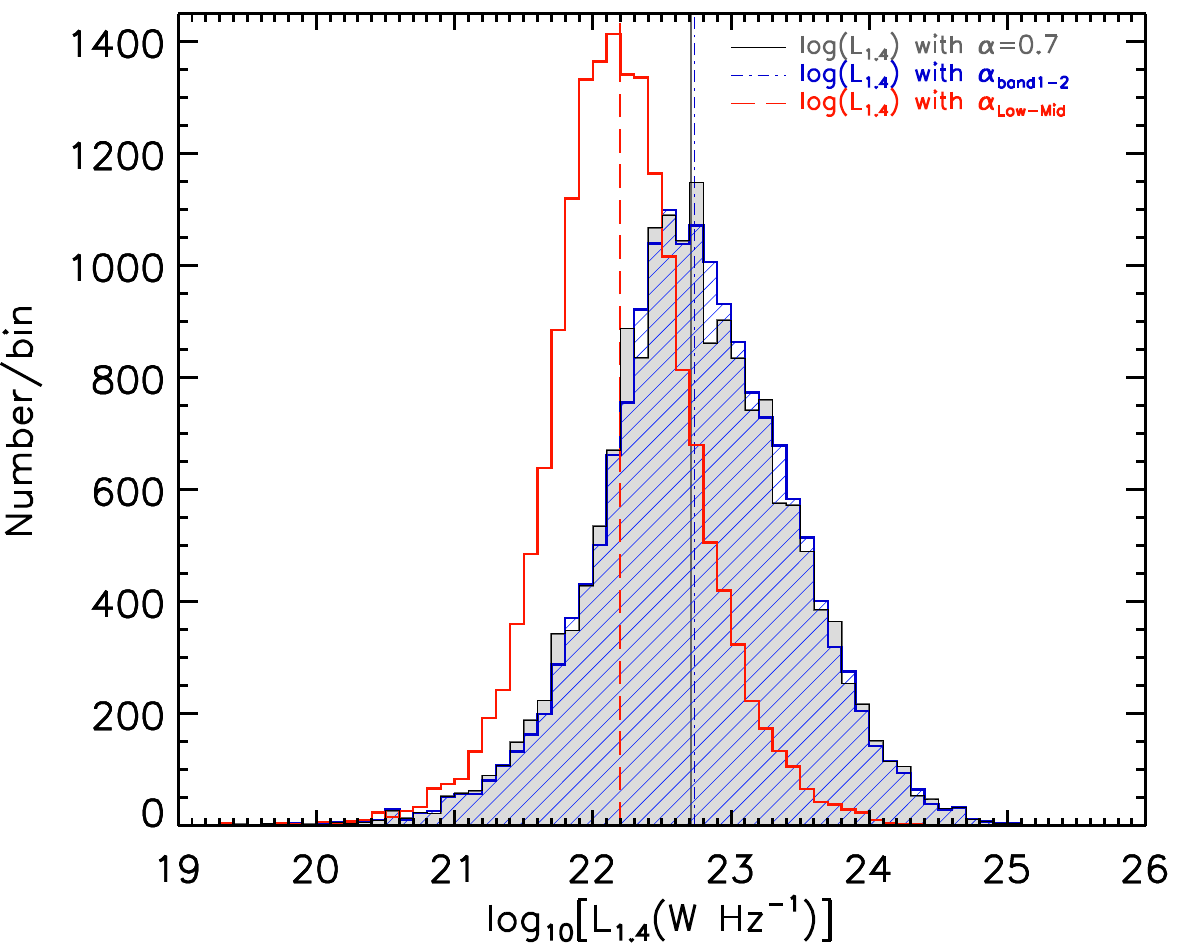}
    \hfill
    \includegraphics[width=0.49\linewidth]{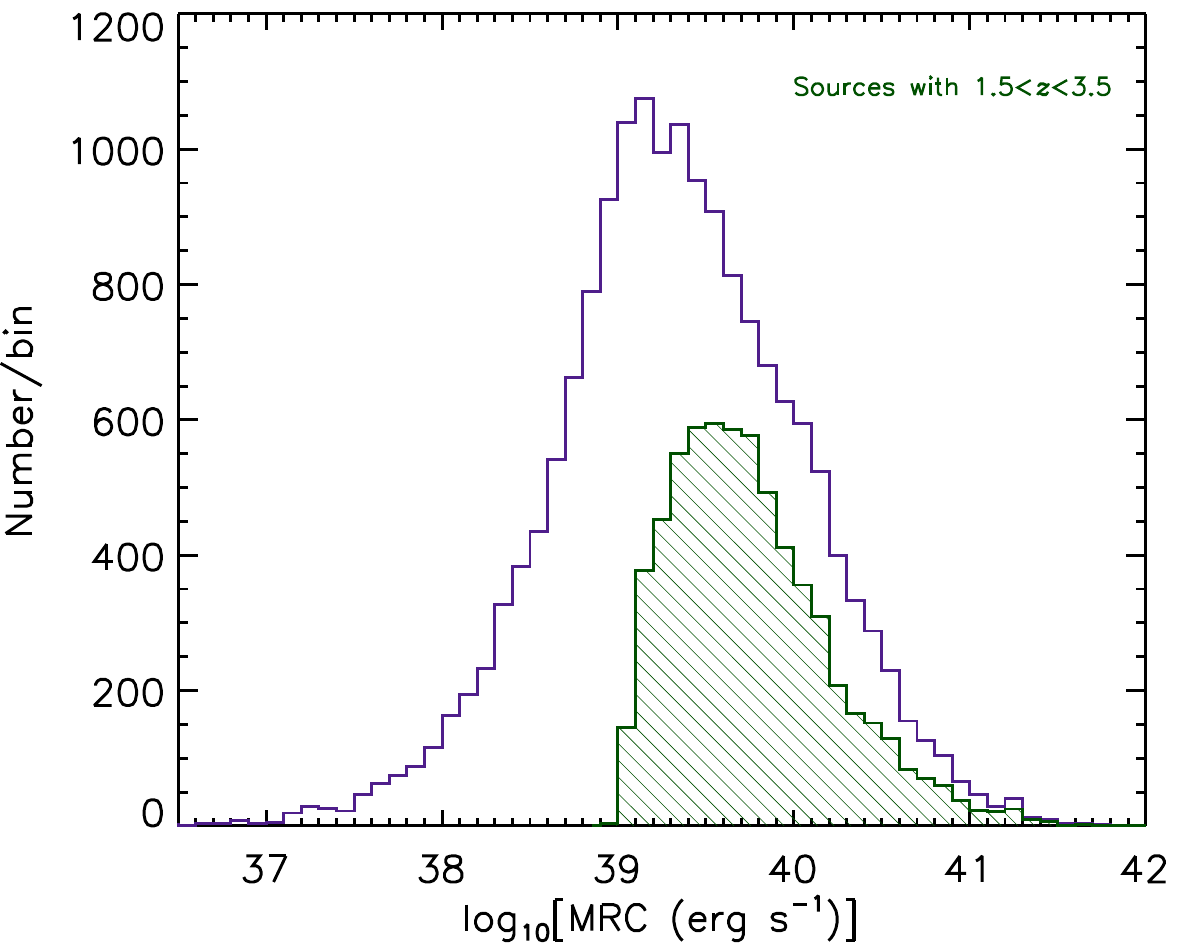}
    \caption{{\it Left:} Distributions of rest-frame 1.4\,GHz luminosities derived using three spectral-index assumptions in the $k$-correction: a fixed $\alpha=0.70$ (filled grey), a two-point spectral index measured between SKA-Mid Bands\,1 and 2 (blue hatched), and a spectral index fitted using the {T-RECS tabulated flux densities from 150\,MHz to 20\,GHz, covering most of the SKA-Low frequency range and the full SKA-Mid range considered here (red). {\it Right:} Distribution of the integrated mid-radio continuum (MRC) luminosity. The green-filled histogram shows the subsample at $1.5<z<3.5$, corresponding to the redshift range of the observational sample discussed in Section~$\S$\ref{sec:SFR calibarion(z)}.}}
    \label{fig:RL_distribution}
\end{figure}

Figure~\ref{fig:RL_distribution} compares the distributions of rest-frame 1.4\,GHz luminosities derived under the three spectral-index assumptions. Assuming a fixed spectral index of $\alpha=0.70$ yields a median luminosity of $L_{1.4\,{\rm GHz}}=5.16\times10^{22}$\,W\,Hz$^{-1}$, with a scatter of $\sigma=0.70$\,dex. When the two-point spectral index measured between SKA-Mid Bands\,1 and 2 is used, the median luminosity increases to $L_{1.4\,{\rm GHz}}=7.59\times10^{22}$\,W\,Hz$^{-1}$, with a comparable scatter of $\sigma=0.69$\,dex. The difference between these two estimates is modest, mainly because the Band\,1--Band\,2 spectral indices are narrowly distributed, with a median value of $\alpha_{\rm Band\,1-2}=0.76$ and a scatter of $\sigma=0.08$.

By contrast, when the spectral index derived from the full SKA-Low to SKA-Mid frequency coverage is used for the $k$-correction, the median luminosity decreases to $L_{1.4\,{\rm GHz}}=1.57\times10^{22}$\,W\,Hz$^{-1}$, with a smaller scatter of $\sigma=0.54$\,dex. This reduction reflects the fact that the radio spectra of SFGs are not necessarily well described by a single power law over a broad frequency range. In particular, accounting for the low-frequency spectral flattening reported by \citet{An21,An24} results in systematically lower inferred rest-frame 1.4\,GHz luminosities compared to estimates based on a fixed spectral index of $\alpha=0.70$.

This effect provides a plausible explanation for why the radio-derived cosmic SFRD reported for the MeerKAT DEEP2 field by \citet{Matthews24}, where a fixed spectral index of $\alpha=0.70$ was assumed, lies systematically above UV--IR-based cosmic SFRD estimates. Recent MIGHTEE-based work further supports this interpretation, showing that more recent radio--SFR relations, based on radio luminosities derived from full SED modelling, can resolve this apparent discrepancy (\citealt{Thykkathu26} and Tabatabaei, Khademi et al. in prep.).

{The right panel of Figure~\ref{fig:RL_distribution} shows the distribution of integrated rest-frame MRC (1--10\,GHz) luminosities derived from the best-fitting radio SEDs. The green-filled histogram highlights the subsample at $1.5<z<3.5$, corresponding to the redshift range of the observational sample discussed in Section~$\S$\ref{sec:SFR calibarion(z)}. Owing to the greater depth of the mock sample, the MRC distribution extends to significantly lower luminosities, reaching $\sim 10^{39}$\,erg\,s$^{-1}$, whereas its bright end is broadly consistent with that of the observed sample presented by \citet{taba_2025}. This demonstrates that SKA AA4 will enable radio-SFR studies to substantially fainter galaxy populations than are currently accessible.}

Overall, the above analysis highlights the critical role of {SKA AA4's broad frequency coverage} in characterising the radio spectra of SFGs. {By enabling direct measurements of spectral curvature and low-frequency flattening, broad-band radio SED fitting provides more reliable $k$-corrections and therefore more accurate estimates of both rest-frame monochromatic radio luminosities and integrated MRC luminosities. These improvements are essential for establishing robust radio--SFR calibrations (Section~$\S$\ref{sec:calibration}) and for obtaining better-constrained radio-based measurements of the cosmic SFH.}

\subsubsection{The Impact of SKA Broad Frequency Coverage on RLF Constraints}
{Having shown that broad frequency coverage can significantly affect the inferred rest-frame radio luminosities, we now examine how these differences propagate into measurements of the RLF. We focus on the Deep multi-band Tier, which provides the broad SKA-Low to SKA-Mid frequency coverage required for source-by-source radio SED fitting.

\begin{figure}
    \centering
    \includegraphics[width=0.8\linewidth]{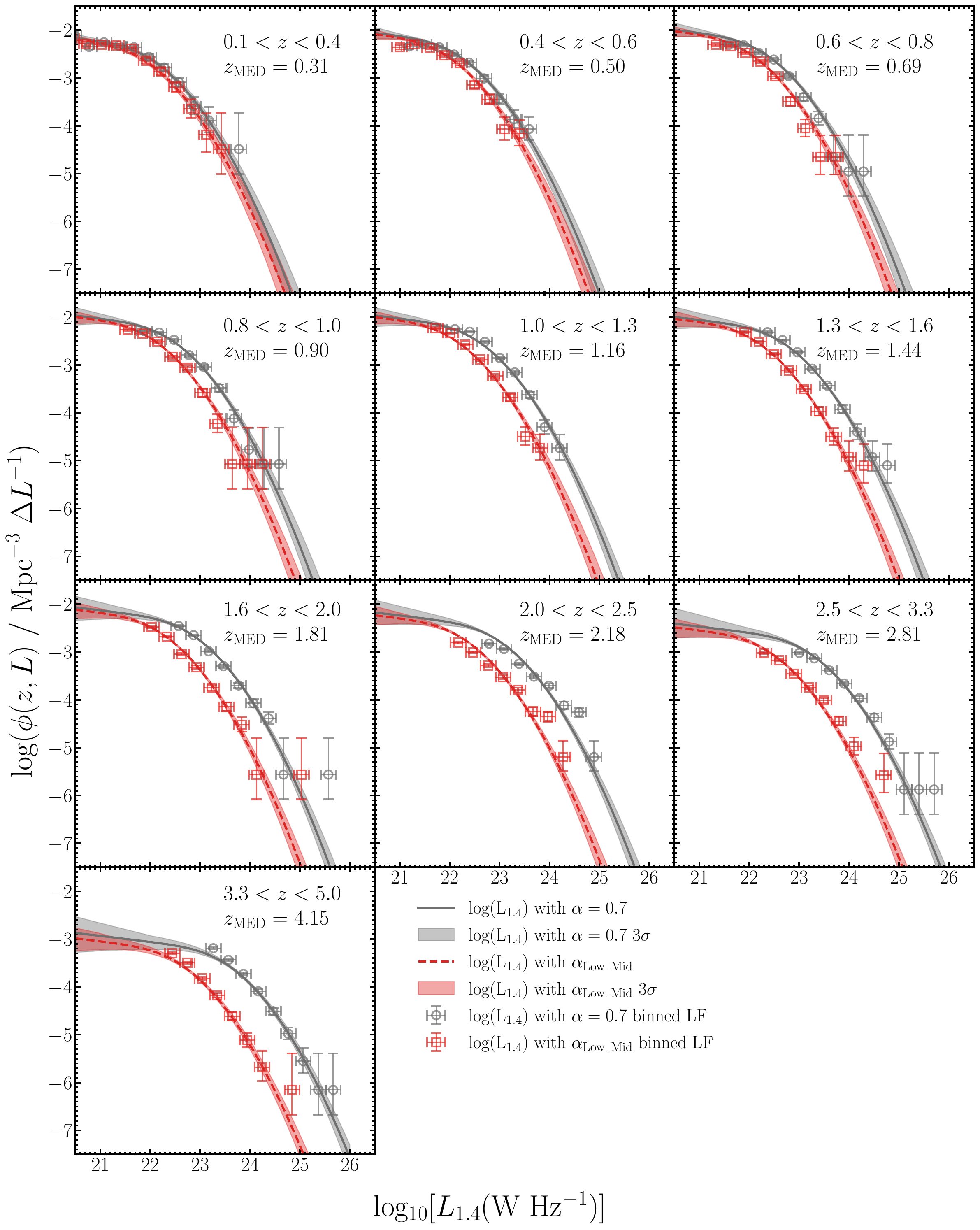}
\caption{Rest-frame 1.4 GHz RLFs in different redshift bins. 
The points with error bars show the binned RLFs estimated using the method of \citet{Page2000}, while the smooth curves are derived from a full maximum-likelihood analysis of the underlying sample. 
Grey symbols and curves correspond to luminosities computed assuming a fixed spectral index of $\alpha=0.70$, appropriate when only SKA-Mid Band 2 observations are available. Red symbols and curves correspond to luminosities computed using the best-fitting spectral index obtained from Bayesian radio-SED fitting over the {broad} SKA-Low to SKA-Mid frequency coverage. The shaded regions indicate the corresponding $3\sigma$ confidence intervals.}
\label{fig:RLF}
\end{figure}

As shown in Figure~\ref{fig:RL_distribution},} the rest-frame 1.4\,GHz luminosities derived using a fixed spectral index of $\alpha=0.70$ and those based on the SKA-Mid Band\,1--2 two-point spectral index show only modest statistical differences. {We therefore} fit the RLFs using two representative luminosity estimates: $L_{1.4\,{\rm GHz}}$ computed with a fixed spectral index ($\alpha=0.7$), and $L_{1.4\,{\rm GHz}}$ computed with the best-fitting spectral index from the {broad} SKA-Low to SKA-Mid radio SED, denoted as $\alpha_{\rm Low-Mid}$.

For each of these two luminosity estimates, we first derive binned RLFs following the method of \citet{Page2000}; these non-parametric measurements are shown as the points with error bars in Figure~\ref{fig:RLF}. We then fit the full unbinned sample using a maximum-likelihood approach, with the evolving RLF parameterised by the LADE model defined in Equation~(\ref{eq:RLF}). The resulting best-fitting models are shown as the smooth curves in Figure~\ref{fig:RLF}.

We explore a broad set of commonly used functional forms for the density evolution, $g(z)$, and luminosity evolution, $f(z)$. For each of the two luminosity estimates, the preferred combination is selected using both the Akaike information criterion (AIC) and the Bayesian information criterion (BIC). Among the 25 combinations tested, both criteria favour the same form:
\begin{equation}
\label{eq:evolution}
g(z)=(1+z)^{p_1+p_2\ln(1+z)}, 
\qquad
f(z)=(1+z)^{k_1+k_2 z}.
\end{equation}
The parameters $p_1$, $p_2$, $k_1$, and $k_2$, together with those describing the local RLF, $\Phi_0(L)$, are then estimated using MCMC.

The resulting best-fitting RLFs are shown as the smooth curves in Figure~\ref{fig:RLF}. They agree well with the binned RLF estimates over the full redshift range. Since the binned estimates and maximum-likelihood fits are obtained independently, this consistency provides a useful cross-check on the inferred RLF evolution.

It is also clear from Figure~\ref{fig:RLF} that the two luminosity estimates yield noticeably different RLFs, particularly at high redshift, where the offset can reach $\sim1$\,dex. This systematic difference indicates that adopting a fixed spectral index bias the inferred RLF evolution. It therefore highlights the importance of deriving source-by-source spectral indices through broad-band radio SED fitting, rather than relying solely on a fixed spectral-index assumption.

\subsection{Sensitivity and Survey Power of SKA AA4 in Constraining the Cosmic SFH}\label{sec:sensitivity_SKA}

The preceding subsection demonstrates that the broad frequency coverage of SKA-Low and SKA-Mid is essential for constructing well-constrained radio SEDs of SFGs {at different redshifts}, which in turn enables reliable $k$-corrections and robust estimates of rest-frame radio luminosities. However, to fully exploit this capability for constraining the cosmic SFH, the sensitivity, survey speed, and field of view of SKA AA4 are equally important. Together, these instrumental properties determine how deeply and how widely the radio sky can be surveyed, directly affecting our ability to sample galaxies across luminosity and redshift and to derive statistically robust measurements of the RLF and its evolution.


\subsubsection{Constraining RLFs with the Ultra-deep Tier}
To assess {the impact of survey sensitivity on SKA AA4 constraints on} the cosmic SFH, we apply {the Ultra-deep}
survey parameters from \cite{Prandoni01.2026.SKA} to the T-RECS mock SFG catalogue \citep{Bonaldi19}. {This survey yields 258,878 mock sources over 1\,deg$^{2}$, selected at a $5\sigma$ limit of $S_{1.4\,{\rm GHz}}>0.25\,\mu$Jy (right panel of Figure~\ref{fig:Mock_SFG}). This setup represents the deepest survey strategy considered here, providing access to faint and high-redshift SFG populations.}

\begin{figure}
    \centering
    \includegraphics[width=0.8\linewidth]{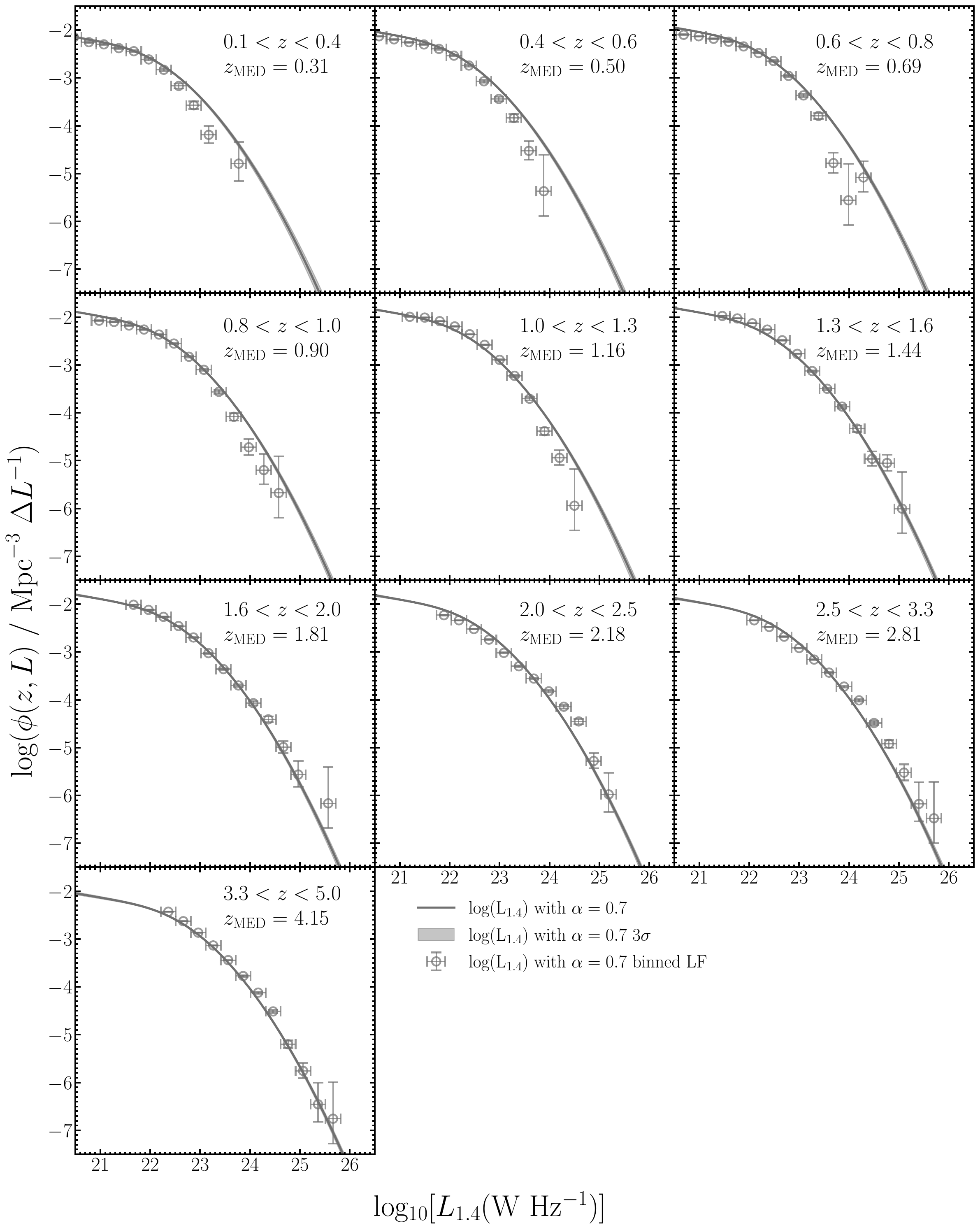}
\caption{Rest-frame 1.4\,GHz RLFs in different redshift bins derived from the T-RECS mock SFG catalogue after applying the Ultra-deep survey parameters from \cite{Prandoni01.2026.SKA}. The RLFs are estimated using the same method as in Figure~\ref{fig:RLF}, but with rest-frame 1.4\,GHz luminosities derived assuming a fixed spectral index of $\alpha=0.70$.}
\label{fig:RLF_ultra_deep}
\end{figure}

{For the Ultra-deep Tier, matched-sensitivity multi-band coverage across the full SKA frequency range is not achievable owing to source-confusion limits at low frequencies. We therefore assume a fixed radio spectral index of $\alpha=0.70$ when estimating rest-frame 1.4\,GHz luminosities and fitting the RLFs.

As shown in Figure~\ref{fig:RLF_ultra_deep}, the statistical uncertainties are substantially reduced compared to those obtained from the Deep multi-band Tier (Figure~\ref{fig:RLF}), primarily owing to the much larger sample size. This improvement is particularly significant for the faint end of the RLF in each redshift bin, and especially at the highest redshifts. At $z\sim5$, the Deep multi-band Tier is limited to detecting relatively luminous starburst galaxies with SFRs of $\gtrsim$100\,M$_{\rm \odot}\,{\rm yr}^{-1}$, and therefore mainly probes the bright luminosity regime above the RLF knee. In contrast, the Ultra-deep Tier can reach main-sequence-like SFGs with SFRs of $\lesssim1\,M_{\odot}\,{\rm yr}^{-1}$ at $z\sim5$, providing much stronger constraints on the faint end of the RLF in the highest-redshift bin. The increased sensitivity therefore greatly improves the statistical reliability of high-redshift faint-end RLF measurements.}

\subsubsection{Forecasting SKA AA4 Constraints on the Cosmic SFH}

Figure~\ref{fig:RLF} and {~\ref{fig:RLF_ultra_deep} demonstrate the potential of SKA AA4 surveys to constrain the cosmic SFH through improved measurements of the evolving RLF. The Deep multi-band Tier illustrates the importance of broad frequency coverage for obtaining well-constrained radio SEDs, reliable $k$-corrections, and robust rest-frame radio luminosities, while the Ultra-deep Tier highlights the impact of increased sensitivity on recovering faint and high-redshift SFG populations. Together, these improvements directly affect the accuracy and statistical robustness of radio-based cosmic SFH estimates.}

This represents a substantial improvement over current radio-continuum constraints. For example, the recent MeerKAT-based cosmic SFH analysis by \citet{Thykkathu26} reaches $z\sim4.5$, but with a flux-density limit of $\sim40\,\mu$Jy. {This limit is significantly shallower than the $5\sigma$ limits adopted here for the SKA AA4 Deep multi-band Tier ($2\,\mu$Jy) and Ultra-deep Tier ($0.25\,\mu$Jy).} As a result, the highest-redshift bins in \citet{Thykkathu26} ($z>2.5$) primarily sample the luminous end of the RLF, with limited leverage on the faint-end population. In contrast, the much greater depth of SKA AA4 will enable direct constraints on the RLF over a broader luminosity range at high redshift, {including sources below the characteristic luminosity, or ``knee'', of the RLF. This is critical for reducing the extrapolation required when integrating the RLF to derive the cosmic SFRD.

Beyond broad frequency coverage and high sensitivity,} the wide field of view (SKA-Low: 2.3–113\,deg$^{2}$; SKA-Mid: 0.007–12.5\,deg$^{2}$) and high survey speed of SKA AA4 will enable the construction of statistically robust samples of SFGs across cosmic time. Combined with the array’s high angular resolution, {these capabilities} will facilitate more accurate and efficient identification of optical/IR counterparts, as well as reliable classification of radio sources into SF-dominated or AGN-dominated populations. 

{Therefore, constraining the cosmic SFH with SKA AA4 will require a combination of complementary survey tiers, as described by \citet{Prandoni01.2026.SKA}. Such a tiered strategy combines broad frequency coverage, high sensitivity, and different survey areas, thereby balancing the need for accurate radio SED characterisation, faint-source detection, and statistically representative samples. These combined capabilities will} (1) improve the accuracy of radio luminosity estimates and radio--SFR calibrations (Section~$\S$\ref{sec:calibration}) through well-constrained radio SEDs; (2) mitigate the effects of cosmic variance and small-number statistics, which currently limit deep but narrow radio surveys; and (3) enable the detection of SFGs below the characteristic luminosity of the RLF at high redshift (Figure~\ref{fig:RLF_ultra_deep}). {Together, these advances will improve constraints on the faint-end slope and redshift evolution of the RLF, thereby providing more robust radio-based measurements of the cosmic SFH.}

\subsection{Constraining Cosmic SFH with SKA AA4: Key Challenges}\label{sec:Challenges_SKA}

\begin{figure}
\centering
\includegraphics[width=0.6\columnwidth]{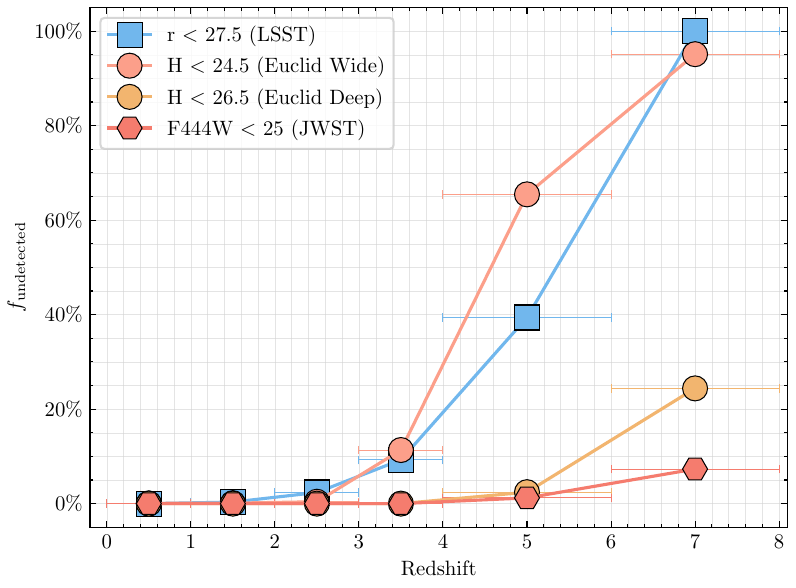}
\caption{Predicted fraction of SKA-detected sources without optical counterparts from LSST. Blue squares, yellow circles and red hexagons denote respectively the fraction of SKA detected galaxies {($S_{1.4\,\rm GHz}\ge2\,\mu$Jy, corresponding to the $5\sigma$ detection threshold of the Deep multi-band Tier) that are fainter than the limiting magnitudes of the LSST $r$ band ($r=27.5$), the Euclid Wide and Deep surveys ($H=24.5$ and 26.5, respectively), and the JWST F444W band ($F444W=25$) over $0<z<8$. The figure} illustrates that deep mid-infrared imaging is essential to detect faint SKA sources at $z > 4$. Figure adopted and modified from \cite{GaoF2025}.
}
\label{fig:SKA_counterparts}
\end{figure}

Although we have outlined the unprecedented capabilities of SKA AA4, including its broad frequency coverage, high sensitivity, high angular resolution, and wide field of view, these attributes also give rise to new observational and analytical challenges. 

\begin{itemize}
    \item Potential lack of optical/near-infrared counterparts at $z>4$, even with deep surveys such as LSST and \textit{Euclid} (see \citealt{GaoF2025} and Figure~\ref{fig:SKA_counterparts}). The extremely high sensitivity of SKA AA4 { is expected to enable}, for the first time, {a more complete census of star formation} in massive galaxies at $z \sim 4-8$ (Figure~\ref{fig:Mock_SFG}). { However, a significant fraction} of these massive galaxies have been shown to be relatively faint, or completely undetected, in the observed optical to near-infrared wavelengths {owing to dust obscuration}~\citep{WangT2016,Wang19,Talia21}. This could hamper the determination of the source redshift (one of several key measurements). These elusive, heavily dust-obscured systems therefore constitute a significant and previously under-appreciated component of cosmic {SFR, particularly at high redshift. As shown in Figure~\ref{fig:SKA_counterparts}, deep mid-infrared imaging would be required to identify most, if not all, of the high-$z$ counterparts of SKA sources. In addition to enabling redshift determination, mid-infrared data are} also essential for constraining the physical properties of these high-z galaxies, including their stellar masses~\citep{WangT2025} and star formation status~\citep{YangT2026}.
    \item Possible contamination from AGNs, especially radio-quiet ones \citep[e.g., see][and references therein]{Panessa2019,WangY2024}, which may complicate the identification of pure SFGs. This issue is further exacerbated at high redshift, where a significant fraction of SKA-detected radio sources lack optical/NIR counterparts. {Multi-wavelength diagnostics, such as machine learning (e.g., \citealt{Silima25,Mazoochi}) and spectral energy distribution fitting (e.g., \citealt{Azadi23}), are therefore needed to reliably disentangle AGN- and SF-dominated radio emission.}  
    \item {At high redshift, the increased gas density within galaxies may lead to greater uncertainties in their radio spectral shapes \citep[e.g.,][]{grundy:25}. {This} further emphasizes the necessity for {broad multi-frequency coverage to decompose the radio spectrum and obtain a more reliable} SFR.}
    \item {As discussed in Section~$\S$\ref{sec:calibration}, the MRC--SFR relation appears tighter and closer to linear than calibrations based on monochromatic radio luminosities, such as {the} $L_{\rm 1.4\,GHz}$--SFR. This makes the MRC a promising tracer of star formation in individual galaxies. However, using MRC luminosities to trace the cosmic SFH requires a robust LF framework for integrated radio luminosities. In our analysis, we find that LF parameterisations commonly used for monochromatic radio luminosities do not directly yield stable results when applied to the integrated MRC. This is expected because the MRC {is an integrated radio luminosity and depends on the rest-frame 1--10\,GHz radio SED shape, rather than on the luminosity at a single reference frequency.} Developing an LF framework specifically suited to integrated MRC luminosities, {together with a consistent method for converting such LFs into cosmic SFRD measurements,} will therefore be an important topic for future work.}
\end{itemize}

\section*{Acknowledgement}
FXA acknowledges the support from the National Natural Science Foundation of China (12303016), the Natural Science Foundation of Jiangsu Province (BK20242115), and Yunnan Revitalization Talent Support Program Young Talent Project. FST acknowledges the support from the DYNAVERSE Cluster of Excelle (Cologne – Bonn) which is
Funded by the Deutsche Forschungsgemeinschaft (DFG, German Research Foundation) under Germany's Excellence Strategy – EXC 3037 – 533607693. This research was supported by the Commonwealth through an Australian Government Research Training Program Scholarship [DOI: \url{https://doi.org/10.82133/C42F-K220}]. ZY acknowledges financial support from the Science Fund for Distinguished Young Scholars of Hunan Province (Grant No. 2024JJ2040), and the Major Basic Research Project of Hunan Province (Grant No. 2024JC0001). TW acknowledges the supported by National Natural Science Foundation of China (Grant No.12525302 and 12141301), Basic Research Program of Jiangsu (Grant No. BK20250001), and the Fundamental Research Funds for the Central Universities with Grant no.KG202502. Y.Z. is grateful for support from the National Natural Science Foundation of China (NSFC) under grant No. 12173079. MV acknowledges financial support from the Inter-University Institute for Data Intensive Astronomy (IDIA), a partnership of the University of Cape Town, the University of Pretoria and the University of the Western Cape, and from the South African Department of Science and Innovation's National Research Foundation under the ISARP RADIOMAP Joint Research Scheme (DSI-NRF Grant Number 150551) and the CPRR HIPPO Project (DSI-NRF Grant Number SRUG22031677).

\bibliographystyle{abbrvnat-maxbibnames4}
\input{journal-names}
\bibliography{chapter} 

\end{document}

%% file: journal-names.tex
\newcommand{\actaa}{Acta Astron.} 
\newcommand{\araa}{ARA\&A} 
\newcommand{\aar}{A\&ARv} 
\newcommand{\aapr}{A\&ARv} 
\newcommand{\ab}{Astrobiol.} 
\newcommand{\aj}{AJ} 
\newcommand{\apj}{ApJ} 
\newcommand{\apjl}{ApJL} 
\newcommand{\apjs}{ApJSS} 
\newcommand{\ao}{Appl. Opt.} 
\newcommand{\apss}{Astro. \& Space Sci.} 
\newcommand{\aap}{A\&A} 
\newcommand{\aaps}{A\&AS.} 
\newcommand{\baas}{Bull. Am. Astron. Soc.} 
\newcommand{\caa}{Chinese A\&A} 
\newcommand{\cjaa}{Chinese J. A\&A} 
\newcommand{\cqg}{Class. Quantum Gravity} 
\newcommand{\gal}{Galaxies} 
\newcommand{\gca}{Geo. Cosmo. Acta} 
\newcommand{\icarus}{Icarus} 
\newcommand{\jcap}{JCAP} 
\newcommand{\jgr}{J. Geophys. Res.} 
\newcommand{\jgrp}{J. Geophys. Res. Planets} 
\newcommand{\jqsrt}{J. Quant. Spectrosc. Radiat. Transf.} 
\newcommand{\memsai}{Mem. SAIt} 
\newcommand{\mnras}{MNRAS} 
\newcommand{\nat}{Nature} 
\newcommand{\nastro}{Nat. Astron.} 
\newcommand{\ncomms}{Nat. Commun.} 
\newcommand{\nphys}{Nat. Phys.} 
\newcommand{\na}{New Astron.} 
\newcommand{\nar}{New Astron. Rev.} 
\newcommand{\physrep}{Phys. Rep.} 
\newcommand{\pra}{Phys. Rev. A} 
\newcommand{\prb}{Phys. Rev. B} 
\newcommand{\prc}{Phys. Rev. C} 
\newcommand{\prd}{Phys. Rev. D} 
\newcommand{\pre}{Phys. Rev. E} 
\newcommand{\prx}{Phys. Rev. X} 
\newcommand{\prl}{Phys. Rev. Let.} 
\newcommand{\psj}{Planet. Sci. J.} 
\newcommand{\planss}{Planet. Space Sci.} 
\newcommand{\pnas}{Proc. Natl Acad. Sci. USA} 
\newcommand{\procspie}{Proc. SPIE} 
\newcommand{\pasa}{PASA} 
\newcommand{\pasj}{PASJ} 
\newcommand{\pasp}{PASP} 
\newcommand{\rmxaa}{RMXAA} 
\newcommand{\sci}{Science} 
\newcommand{\sciadv}{Sci. Adv.} 
\newcommand{\solphys}{Sol. Phys.} 
\newcommand{\sovast}{Soviet Ast.} 
\newcommand{\ssr}{Space Sci. Rev.} 
\newcommand{\uni}{Universe} 